\documentclass[twoside,twocolumn,9pt]{article}
\usepackage{extsizes}
\usepackage[super,sort&compress,comma]{natbib} 
\usepackage[version=3]{mhchem}
\usepackage[left=1.5cm, right=1.5cm, top=1.785cm, bottom=2.0cm]{geometry}
\usepackage{balance}
\usepackage{times,mathptmx}
\usepackage{sectsty}
\usepackage{graphicx} 
\usepackage{lastpage}
\usepackage[format=plain,justification=justified,singlelinecheck=false,font={stretch=1.125,small,sf},labelfont=bf,labelsep=space]{caption}
\usepackage{float}
\usepackage{fancyhdr}
\usepackage{fnpos}
\usepackage[english]{babel}
\addto{\captionsenglish}{%
  
}
\usepackage{array}
\usepackage{droidsans}
\usepackage{charter}
\usepackage[T1]{fontenc}
\usepackage[usenames,dvipsnames]{xcolor}
\usepackage{setspace}
\usepackage[compact]{titlesec}
\usepackage{hyperref}

\usepackage{amsmath,amsfonts,amssymb}

\usepackage{epstopdf}

\definecolor{cream}{RGB}{222,217,201}

\begin{document}

\pagestyle{fancy}
\thispagestyle{plain}
\fancypagestyle{plain}{

\fancyhead[C]{\includegraphics[width=18.5cm]{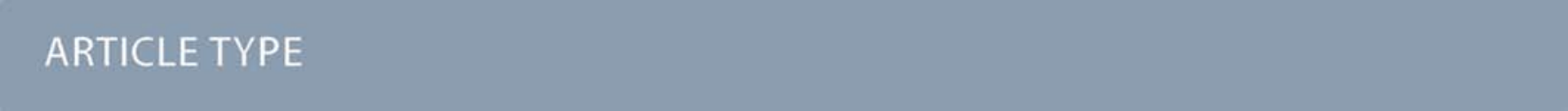}}
\fancyhead[L]{\hspace{0cm}\vspace{1.5cm}\includegraphics[height=30pt]{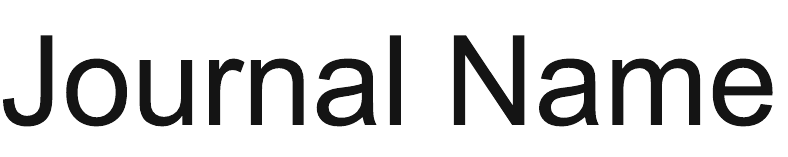}}
\fancyhead[R]{\hspace{0cm}\vspace{1.7cm}\includegraphics[height=55pt]{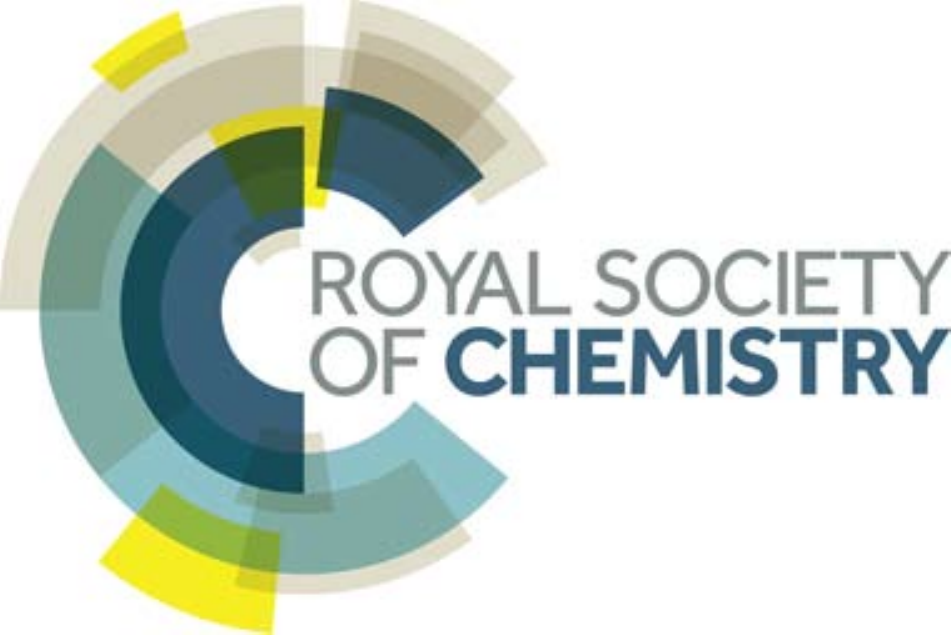}}
\renewcommand{\headrulewidth}{0pt}
}

\makeFNbottom
\makeatletter
\renewcommand\LARGE{\@setfontsize\LARGE{15pt}{17}}
\renewcommand\Large{\@setfontsize\Large{12pt}{14}}
\renewcommand\large{\@setfontsize\large{10pt}{12}}
\renewcommand\footnotesize{\@setfontsize\footnotesize{7pt}{10}}
\makeatother

\renewcommand{\thefootnote}{\fnsymbol{footnote}}
\renewcommand\footnoterule{\vspace*{1pt}%
\color{cream}\hrule width 3.5in height 0.4pt \color{black}\vspace*{5pt}} 
\setcounter{secnumdepth}{5}

\makeatletter 
\renewcommand\@biblabel[1]{#1}            
\renewcommand\@makefntext[1]%
{\noindent\makebox[0pt][r]{\@thefnmark\,}#1}
\makeatother 
\renewcommand{\figurename}{\small{Fig.}~}
\sectionfont{\sffamily\Large}
\subsectionfont{\normalsize}
\subsubsectionfont{\bf}
\setstretch{1.125} 
\setlength{\skip\footins}{0.8cm}
\setlength{\footnotesep}{0.25cm}
\setlength{\jot}{10pt}
\titlespacing*{\section}{0pt}{4pt}{4pt}
\titlespacing*{\subsection}{0pt}{15pt}{1pt}

\fancyfoot{}
\fancyfoot[LO,RE]{\vspace{-7.1pt}\includegraphics[height=9pt]{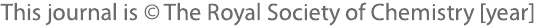}}
\fancyfoot[CO]{\vspace{-7.1pt}\hspace{13.2cm}\includegraphics{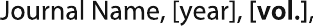}}
\fancyfoot[CE]{\vspace{-7.2pt}\hspace{-14.2cm}\includegraphics{head_foot/RF}}
\fancyfoot[RO]{\footnotesize{\sffamily{1--\pageref{LastPage} ~\textbar  \hspace{2pt}\thepage}}}
\fancyfoot[LE]{\footnotesize{\sffamily{\thepage~\textbar\hspace{3.45cm} 1--\pageref{LastPage}}}}
\fancyhead{}
\renewcommand{\headrulewidth}{0pt} 
\renewcommand{\footrulewidth}{0pt}
\setlength{\arrayrulewidth}{1pt}
\setlength{\columnsep}{6.5mm}
\setlength\bibsep{1pt}

\makeatletter 
\newlength{\figrulesep} 
\setlength{\figrulesep}{0.5\textfloatsep} 

\newcommand{\topfigrule}{\vspace*{-1pt}%
\noindent{\color{cream}\rule[-\figrulesep]{\columnwidth}{1.5pt}} }

\newcommand{\botfigrule}{\vspace*{-2pt}%
\noindent{\color{cream}\rule[\figrulesep]{\columnwidth}{1.5pt}} }

\newcommand{\dblfigrule}{\vspace*{-1pt}%
\noindent{\color{cream}\rule[-\figrulesep]{\textwidth}{1.5pt}} }

\makeatother

\graphicspath{{./FIGURES/}}


\let\a=\alpha \let\b=\beta \let\g=\gamma \let\d=\delta
\let\e=\epsilon \let\z=\zeta \let\h=\eta \let\k=\kappa
\let\l=\lambda \let\m=\mu \let\n=\nu \let\x=\xi \let\p=\pi
\let\s=\sigma \let\t=\tau \let\f=\varphi \let\ph=\varphi\let\c=\chi
\let\ps=\psi \let\y=\upsilon \let\si=\varsigma \let\G=\Gamma
\let\D=\Delta \let\Th=\Theta\let\L=\Lambda \let\X=\Xi \let\P=\Pi
\let\Si=\Sigma \let\F=\Phi \let\Ps=\Psi \let\Y=\Upsilon
\let\ee=\varepsilon \let\r=\rho \let\th=\theta \let\io=\infty
\let\om=\omega
\def\ie{{i.e. }}\def\eg{{e.g. }}

\def\PP{{\cal P}}\def\EE{{\cal E}}\def\MM{{\cal M}} \def\VV{{\cal V}}
\def\CC{{\cal C}}\def\FF{{\cal F}} \def\HH{{\cal H}}\def\WW{{\cal W}}
\def\TT{{\cal T}}\def\NN{{\cal N}} \def\BB{{\cal B}}\def\II{{\cal I}}
\def\RR{{\cal R}}\def\LL{{\cal L}} \def\JJ{{\cal J}} \def\OO{{\cal O}}
\def\DD{{\cal D}}\def\AA{{\cal A}}\def\GG{{\cal G}} \def\SS{{\cal S}}
\def\KK{{\cal K}}\def\UU{{\cal U}} \def\QQ{{\cal Q}} \def\XX{{\cal X}}
\def\YY{{\cal Y}}\def\ZZ{{\cal Z}}

\newcommand {\tw}	{t_\mathrm{w}}
\newcommand {\fg}	{\f_{\mathrm{g}}}
\newcommand {\Pg}	{P_{\mathrm{g}}}
\newcommand {\fG}	{\f_{\mathrm{G}}}
\newcommand {\PG}	{P_{\mathrm{G}}}

\newcommand{\beq}{\begin{equation}}
\newcommand{\eeq}{\end{equation}}
\newcommand{\av}[1]{{\left\langle {#1} \right\rangle}}


\twocolumn[
  \begin{@twocolumnfalse}
\vspace{3cm}
\sffamily
\begin{tabular}{m{4.5cm} p{13.5cm} }

\includegraphics{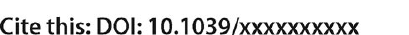} & \noindent\LARGE{\textbf{Spin-glass--like aging in colloidal and granular glasses}} \\
\vspace{0.3cm} & \vspace{0.3cm} \\

 & \noindent\large{Beatriz Seoane$^{\ast}$\textit{$^{a,b}$} and Francesco Zamponi\textit{$^{a}$}} \\

\includegraphics{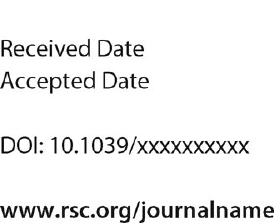} & \noindent\normalsize{
Motivated by the mean field prediction of a Gardner phase transition 
between a ``normal glass'' and a ``marginally stable glass'', we investigate the off-equilibrium dynamics of three-dimensional polydisperse 
hard spheres, used as a model for colloidal or granular glasses.
Deep inside the glass phase,
we find that a sharp crossover pressure $P_{\rm G}$ separates two distinct dynamical regimes. For pressure $P < P_{\rm G}$,
the glass behaves as a normal solid, displaying fast dynamics that quickly equilibrates within the glass free energy basin. For $P>P_{\rm G}$, instead, the dynamics
becomes strongly anomalous, displaying very large equilibration time scales, aging, and a constantly increasing dynamical susceptibility. The crossover at $P_{\rm G}$ is strongly reminiscent of the one
observed in three-dimensional spin-glasses in an external field, suggesting that the two systems could be in the same universality class, 
consistently with theoretical expectations.
} \\

\end{tabular}

 \end{@twocolumnfalse} \vspace{0.6cm}

  ]

\renewcommand*\rmdefault{bch}\normalfont\upshape
\rmfamily
\section*{}
\vspace{-1cm}


\footnotetext{\textit{$^{a}$Laboratoire de physique th\'eorique, D\'epartement de physique de
l'ENS, \'Ecole normale sup\'erieure, PSL
Research University, Sorbonne Universit\'es, CNRS,
75005 Paris, France;}}
\footnotetext{\textit{$^{b}$Instituto de Biocomputaci\'on y F\'{\i}sica de Sistemas Complejos (BIFI), 50009 Zaragoza, Spain. }}
\footnotetext{\textit{$^{\ast}$E-mail: beaseobar@gmail.com}}




\section{Introduction}

The Gardner transition is an exotic spin-glass phase transition that was discovered independently in 1985 by Gardner~\cite{Ga85} and by Gross, Kanter, and Sompolinsky~\cite{GKS85}.
It happens in a broad range of mean field spin-glass models, that belong to the discontinuous 
random first order transition (RFOT) universality class~\cite{De80,GM84,KW87b,KT88,CC05}. 
The physics of these models is the following. At high temperature $T$, there is a single paramagnetic
phase. Upon lowering the temperature, a discontinuous {\it dynamical glass transition} is met at $T=T_{\rm d}$, at which the dynamics 
become arrested, while the equilibrium properties display no singularity.
At any temperature $T<T_{\rm d}$,
a large number of infinitely long-lived metastable glassy states are present. These states fully trap the dynamics and ergodicity is broken, 
but if prepared inside any of them, the system can fully equilibrate in a short time. One can then prepare a system in equilibrium in the arrested phase, 
at an initial temperature $T_{\rm g} < T_{\rm d}$, and change the temperature to a value $T < T_{\rm g}$.
The system remains confined in 
the original glass state, whose evolution can thus be followed in a restricted, metastable equilibrium~\cite{FP95,BBM96,BFP97,KZ10,KZ10b}. 
During this process, each individual state might undergo a continuous spin-glass phase transition at a temperature $T_{\rm G}(T_{\rm g}) < T_{\rm g}$, 
called the Gardner transition~\cite{BFP97,MR04,KZ10,Ri13}. Below the Gardner transition, $T<T_{\rm G}(T_{\rm g})$, the glass state is described by
a full replica symmetry broken (fullRSB) structure, 
discovered by Parisi~\cite{Pa79}, that also describes the low-temperature phase of usual mean field spin-glasses~\cite{MPV87}.

The RFOT universality class is known to provide a mean field theory for the structural glass transition of particle systems~\cite{KW87,KT87,KT87b,KTW89,MP09,Ca09,BB11,WL12}. 
In particular, a $d$-dimensional 
hard sphere liquid in the limit of $d\to\io$ realises precisely the RFOT scenario~\cite{KW87,PZ06a,PZ10,KPUZ13,MKZ16,CKPUZ17}, 
with temperature $T$ being replaced, as control parameter, by packing fraction $\f$ or pressure $P$.
Yet, despite the fact that {\it (i)} the analogy between the structural glass transition and RFOT spin-glass models, and {\it (ii)} the presence of a Gardner transition in the latter models,
are both well known since the 80s, no attempt to look systematically for a Gardner transition in structural glasses has been made, until very recently.

The situation changed dramatically when it was realised that hard sphere glasses, close to the jamming transition~\cite{LN98,OLLN02,OSLN03} 
(i.e. the infinite pressure limit), 
are {\it marginally stable}~\cite{SLN05,WSNW05,Wyart,BW06,BW07,LNSW10,LN10,Wy12,MW15}.
Marginal stability is naturally described, at the mean field level, by the Parisi fullRSB construction~\cite{MPV87}, thus motivating the search for a Gardner transition in structural glasses.
Such a transition was indeed found~\cite{KPUZ13,CKPUZ17} in the limit $d\to\io$. It was then found~\cite{CKPUZ14NatComm,CCPZ15,CKPUZ17} 
that the resulting fullRSB construction in $d\to\io$ provides an accurate 
quantitative description of the criticality and marginality of jamming in any dimension $d\geq 2$, 
leading to the conjecture~\cite{GLN12} that
the lower critical dimension for jamming might be equal to $d=2$.

The existence of a Gardner transition to a fullRSB phase in structural glasses might have deep implications for their properties. 
In fact, marginal stability strongly affects the low-frequency vibrational density of states~\cite{DLDLW14,FPUZ15,MSI17}, 
the elastic~\cite{BU16} 
and rheological~\cite{YZ14,RUYZ15,JY17,UZ17,FS17} properties of the solid, and its low-temperature transport properties~\cite{XVLN10}.
These theoretical results motivated the proposal that marginal stability could be a unifying principle behind many anomalies of amorphous 
solids~\cite{LN10,KPUZ13,DLDLW14,MW15}, and the search for a Gardner transition
in a broad range of systems, both in numerical simulations~\cite{CJPRSZ15,BCJPSZ15,JY17,SBZ17,SRPZ18,charbonneau2018gardner} and 
experiments~\cite{SD16,GLL18}.
Currently, it is believed that $3d$ Lennard-Jones--like systems, that are good models for metallic and molecular glasses, generically do not show a Gardner transition~\cite{SBZ17,SRPZ18}, while $3d$ hard sphere systems, that are good models for granular and colloidal glasses, 
display several anomalies
at high pressures~\cite{CJPRSZ15,BCJPSZ15,JY17}, that suggest the existence of a sharp Gardner crossover, if not a true transition.
However, a careful investigation of the nature of both the crossover and the new phase has not yet been performed.

In this work, we study the nature of the Gardner crossover in $3d$ hard sphere glasses. We report a careful study of the off-equilibrium dynamics
after an instantaneous compression (``crunch'') from the equilibrium, dynamically arrested, supercooled liquid phase, to a state at higher pressure $P$, 
deep in the glass phase. For mild compressions, 
the dynamics relaxes to the metastable glass equilibrium on short times. For compressions to a higher pressure, a long time scale
emerges. The dynamics is unable to ergodically sample the glass basin, it displays aging, and a susceptibility that grows in time. 

We also present results for the $3d$ Edwards-Anderson spin-glass model
with a magnetic field, which display the same qualitative behavior. This comparison
 suggests that the $3d$ spin glasses in a field and $3d$ colloidal
glasses probably fall into the same universality class, as it happens in
$d\to\io$ and as it is generically expected from the theory, because
in both systems there is no spontaneous symmetry breaking apart from
the replica one~\cite{FM13,BU15,BU16,charbonneau2017nontrivial}.  In
both systems, the dynamics becomes extremely slow around a
crossover, due to a cooperative phenomenon associated to a growing
correlation length, but the length grows so slowly (roughly logarithmically) with time that
deciding whether a true divergence (i.e. a phase transition) or just a
sharp dynamical crossover are observed is extremely tricky~\cite{hicks2017gardner}. 
Note that the
analogy between the dynamics of glasses and the $d=3$ Edwards-Anderson
spin-glass in a field had also been previously proposed  in simulations (see
Ref.~\citenum{janus2014dynamical}), although in a different region of
the phase diagram.
Also, note that a RFOT theory of nucleation effects in hard spheres has been recently developed~\cite{LW18}, 
which gives predictions about how marginal stability is modified in three dimensions.

Our results are interesting for the physics of colloidal and granular glasses, because they show that important dynamical anomalies are to be
expected deep in the glass phase, consistently with the experimental results of Ref.~\citenum{SD16}. Moreover, we introduce a set of simple dynamical
observables that can be measured in experiments to reveal the growing length and time scales associated to the Gardner crossover.

\section{Simulation details, preparation protocol, and main observables}
\label{sec:simu}

We simulate a system of $N$ three-dimensional ($3d$) polydisperse hard spheres, precisely identical to the one investigated
in Ref.~\citenum{BCJPSZ15}. In particular, the size polydispersity is chosen to be continuous, where each particle diameter $\sigma$ is taken from a distribution $P(\sigma_{\rm m}\leq\sigma\leq\sigma_{\rm M})=A/\sigma^3$, where $\sigma_{\rm m}/\sigma_{\rm M}=0.4494$. This choice for the polydispersity allows one, at the same time, to prevent crystallization and to equilibrate the system via 
an optimized simulation algorithm (see below) up to unusually high densities~\cite{berthier2016equilibrium}. 
Most of the results are reported for $N=1000$ but we also explored other sizes to
quantify the finite size effects. All the quantities below are given in units of the mean particle diameter (length scale), the temperature (energy scale),
and the particle mass (time scale)~\cite{BCJPSZ15}.

\begin{figure}[t]
\centering
  \includegraphics[width=\columnwidth]{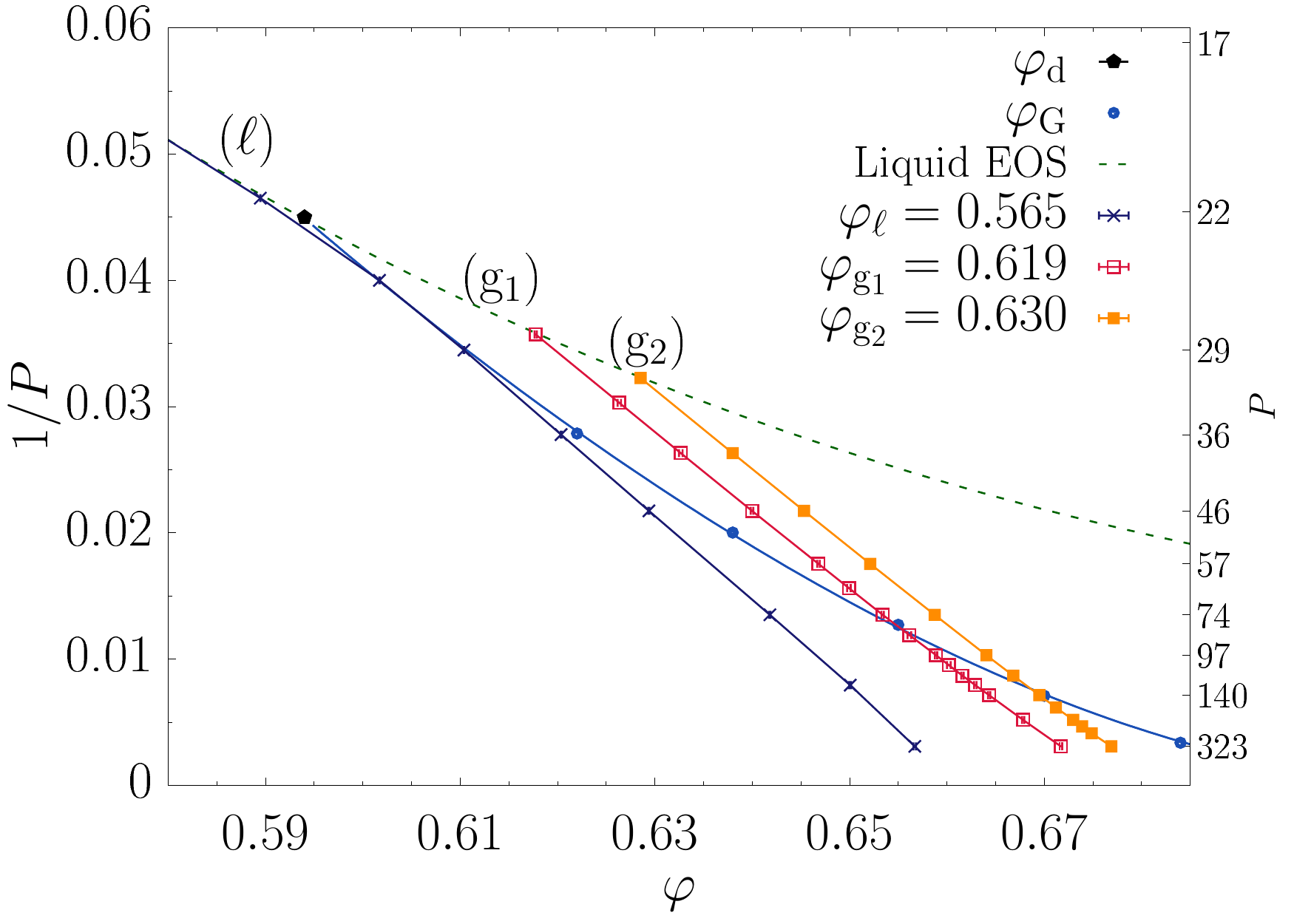}
  \caption{{\bf Phase diagram}. Phase diagram in the inverse pressure $1/P$ vs packing fraction $\f$ plane.
  The dashed line is the supercooled liquid equilibrium equation of state, from Ref.~\citenum{BCJPSZ15}, on which the three initial state points
  are selected at $\f_{\ell}=0.565$, $\f_{{\rm g}_1} = 0.619$ and $\f_{{\rm g}_2} = 0.630$. The black pentagon marks the location of
  the dynamical glass transition $\f_{\rm d}$, also from Ref.~\citenum{BCJPSZ15}. 
  The circles indicate the estimate of the Gardner crossover obtained in Ref.~\citenum{BCJPSZ15}.
  The crosses, open squares, and full squares indicate the off-equilibrium equations of state obtained in this work, 
  by compression from the three initial densities. For later reference, 
  we include the values of pressure $P$ for several of the plotted points on the right-side vertical axis. 
}
  \label{fgr:PD}
\end{figure}

A set of initial configurations at three distinct packing fractions
$\f_{\ell}=0.565$, $\f_{{\rm g}_1} = 0.619$ and $\f_{{\rm g}_2} = 0.630$
(Fig.~\ref{fgr:PD}) have been prepared using an event-driven Molecular
Dynamics (MD) code\footnote{The swap-MD code has been written by
  Yuliang Jin who kindly provided us the initial equilibrated
  configurations, from Ref.~\citenum{JY17}.}, in which in addition to
MD steps, from time to time particles are swapped, allowing for fast
equilibration up to extremely high values of
densities~\cite{berthier2016equilibrium}.  At density $\f_{\ell}=0.565$, the relaxation
time of the standard MD without swap is short compared with the
accessible times, so the system is liquid. At densities $\f_{{\rm g}_1}$
and $\f_{{\rm g}_2}$, the standard relaxation time of the MD dynamics is much larger than the accessible times of the simulation, so the system is in a dynamically
arrested supercooled liquid phase~\cite{berthier2016equilibrium}. 
Note that in the following we will carefully distinguish a ``dynamically
arrested supercooled liquid'', which is an equilibrium liquid whose relaxation time
is much larger than the experimentally accessible time scale, from a ``glass'' which is
a state obtained by compressing off-equilibrium a dynamically arrested system. 
For each initial
density, $N_{\rm s} =20$ (40 for the highest pressures) independent
equilibrium configurations have been prepared (both concerning the
radii and the positions of the particles). We refer to these initial
configurations as our ``samples''.

These equilibrated configurations are used as input to a standard
constant-pressure Monte Carlo (MC) code that mimics the physical evolution of the system: this means that swaps between particles are 
no longer allowed, and only individual local movements of particles, together with global changes of the volume to keep the pressure constant, 
are proposed. Moves are accepted with a standard Metropolis weight.
The target pressure $P$ of
the MC code is set to a larger value than the initial equilibrium
pressure, i.e. the system is instantaneously compressed
(``crunched'')~\cite{MRST02,DAPR00} from the initial equilibrium
pressure to the target pressure $P$. With the objective of fixing the
time-scale and keeping the acceptance rate of the proposed volume
dilation/contraction moves around 0.3 at the final density, we fix $\delta V\propto 1/(P N)$
(note that the probability that two distinct particles overlap grows
with the system size). In addition, we propose single-particle moves
randomly distributed within the cube of side $e^{-10 R}\delta r$,
where $R\in[0,1)$ is a uniformly distributed random variable, and
  $\delta r=1/P$ (to take into account that the cage size
  decreases linearly with the pressure), being $L$ the linear size of
  the system. The random exponential term is introduced to encourage
  relaxation at high pressures, which requires both long and very short
  displacements. Note that with this choice of parameters,
  the bigger the system, the slower the compression of the system is
  after the quench, but once the volume has converged to its final
  value, there is almost no effect of $N$ in the dynamics. We will
  come back to this point in Section~\ref{sec:FS}.

\begin{figure*}[t]
\centering
  \includegraphics[height=7cm]{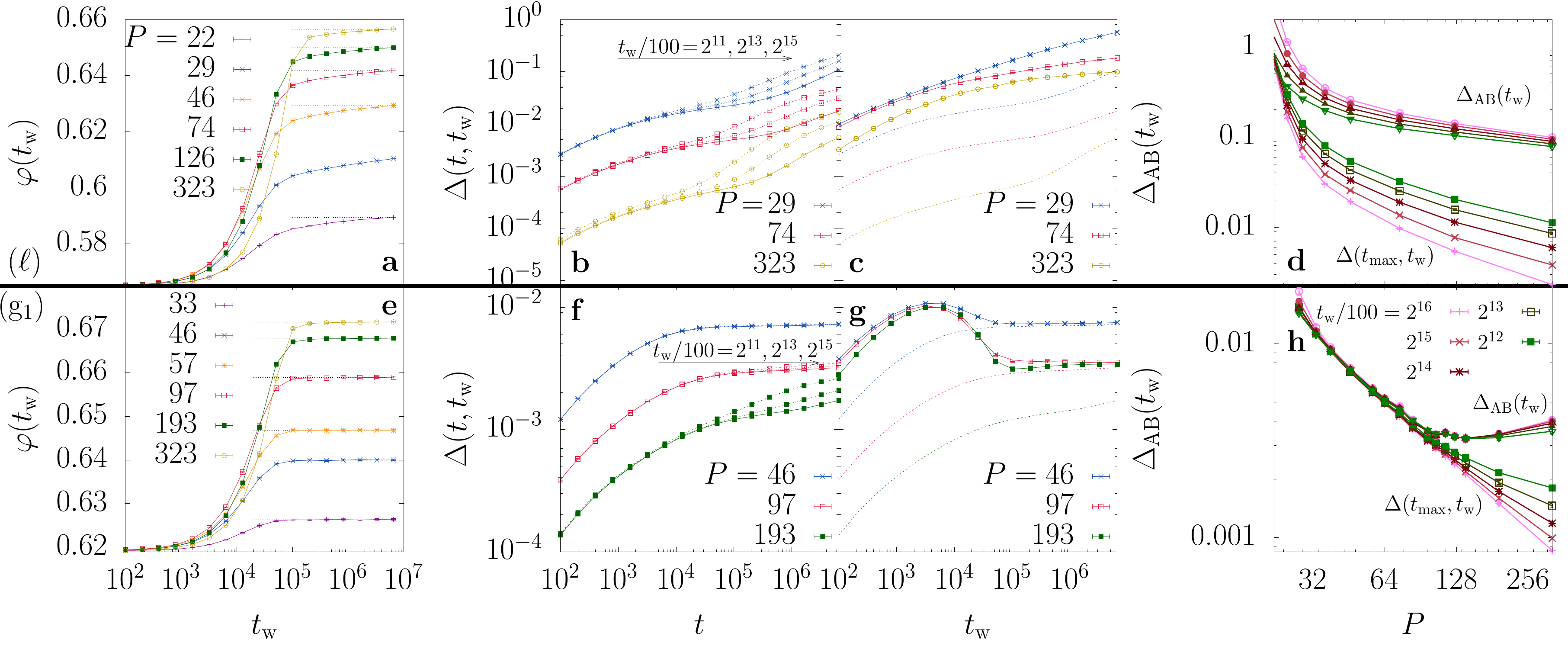}
  \caption{{\bf Dynamics after a crunch.}
  We show the evolution of different dynamical observables
    after a sudden crunch from the ergodic liquid ($\varphi_\ell=0.565$)
    (top panels) and the dynamically arrested liquid ($\varphi_\mathrm{g1}=0.619$)
    (bottom panels). (a,e) {\bf Densification dynamics:} evolution of
    the packing-fraction $\varphi$ with $\tw$, the time elapsed since
    the quench, for different final pressures. While the liquid (a)
    displays extremely slow evolution, the glass (e) quickly converges
    to its final value. As a guide for the eye, we mark the last value
    of $\f(\tw)$ with a dashed horizontal line.  (b,f) {\bf Aging in
      the mean-squared-displacement (MSD):} we show $\Delta(t,\tw)$
    for three final pressures $P$ and three different $\tw$. Again,
    the liquid (b) displays traditional aging behavior for all the
    pressures, i.e. the development of a plateau of caging dynamics,
    followed by diffusion at longer times. The situation in the glass
    is different (f): at low pressures no aging or diffusion is
    observed in the simulated window of times, while strong
    non-diffusive aging is observed for pressures above a well
    defined crossover at $P\sim 97$. (c,d,g,h) {\bf Ergodicity breaking:} Comparison between $\Delta(t,\tw)$ as function of
    $t$ for the highest $\tw$ ($\tw/100=2^{15}$) and the mean-squared
    distance between clones, $\Delta_{\rm AB}(\tw)$ vs $\tw$. Because the
    liquid (c) does not display constrained dynamics at low
    pressures, the different clones are free to separate,
    giving, as a result, large values of $\Delta_{\rm AB}$ even beyond the
    plateau of the MSD. On the contrary, in the glass (g), both
    $\Delta_{\rm AB}$ and $\Delta$ converge to the same plateau value for
    $P\lesssim 97$, evidentiating the thermalization within the glass
    basin, while both values clearly differ at high pressures. In (d)
    and (h) we show (for different values of $\tw$)
    $\Delta(t_\mathrm{max},\tw)$, with $t_\mathrm{max}$ the higher $t$
    available for each $\tw$, and $\Delta_{\rm AB}(\tw)$ as function of
    the pressure $P$. The splitting between $\Delta_{\rm AB}$ and $\Delta$
    at high pressures in the glass (h) suggests a breaking of
    ergodicity beyond the Gardner crossover.
    }
  \label{fgr:dynamics}
\end{figure*}

Following Ref.~\citenum{BCJPSZ15}, each equilibrium configuration is compressed independently (i.e. using a different seed for the random number generator of the MC code) 
several times, leading to a set of $N_{\rm c}=41$ ($81$ for the highest pressures) ``clones'' dynamically evolving at constant target pressure $P$. 
We denote by $\tw$ the time elapsed since the initial crunch,
and we measure the subsequent evolution of several observables as a function of time. 
Besides the instantaneous density $\f(\tw)$,
following previous work~\cite{CJPRSZ15,BCJPSZ15,SBZ17,SRPZ18},
we focus our attention on the mean square
displacement (MSD) of particles in each clone between time $\tw$ and $t+\tw$, defined as
\beq \Delta (t+\tw,\tw) = \frac{1}{N} \sum
_{i=1}^{N} \overline{\av{|\vec{r}_i(t + \tw) - \vec{r}_i(\tw)|^2}} \ ,
\label{eq:MSD}
\eeq
and the MSD between particles in two distinct clones (denoted A and B) 
of the same initial configuration at the same time  $\tw$,
  $\{\vec{r}_i^A(\tw)\}$ and $\{\vec{r}_i^B(\tw)\}$, defined as
\beq \Delta_{\rm AB} (\tw) = \frac{1}{N} \sum
_{i=1}^{N}\overline{ \av{|\vec{r}_i^A(\tw) - \vec{r}_i^B(\tw)|^2} }\ .
\label{eq:DAB}
\eeq In practice, to avoid spurious contributions from small particles
diffusing in the holes of large ones, the sums over $i$ have been
restricted to the $N/2$ largest particles.  Here, $\av{\ \bullet \ }$
refers to the dynamical average, computed as the average over all the
clones of the same sample, while $\overline{\ \bullet \ }$ refers to
the average over all the samples with the same initial density. To
increase the statistics, the dynamical average of $\Delta_{\rm AB}$ is
computed using all the $N_\mathrm{c}(N_\mathrm{c}-1)/2$ possible pairs
of A and B clones, but the error bars are computed by taking into
account the correlations between pairs using the jack-knife
method~\cite{amit2005field}.

In addition, we introduce the displacement of individual particles
$u_i(t,\tw) = |\vec{r}_i(t+\tw) - \vec{r}_i(\tw)|^2$, and following
Ref.~\citenum{SBZ17} we introduce a susceptibility
\begin{eqnarray}\label{eq:chidyndef} 
&\chi(t,\tw) = \overline{\chi^m(t,\tw)}&\\\label{eq:chidyndefsample}
&\text{with  }\chi^m(t,\tw)= \frac{
      \sum_{ij} [ \av{ u_i(t,\tw) u_j(t,\tw)} -
      \av{u_i(t,\tw)}\av{u_j(t,\tw)} ] }{ \sum_i [\av{ u_i(t,\tw)^2} -
      \av{u_i(t,\tw)}^2]}  \ ,&  \end{eqnarray}
  where $\chi^m(t,\tw)$ represents the susceptibility of sample $m$. 
  Note that this quantity is computed only using single clones, which means that the dynamical average is performed over the $N_\mathrm{c}$ clones.
   Similarly, defining
$u^{AB}_i(\tw) = |\vec{r}_i^A(\tw) - \vec{r}_i^B(\tw)|^2$ as the relative
displacement of two clones, we can introduce \begin{eqnarray}\label{eq:chiABdef}
&\chi_{\rm AB}(\tw) = \overline{\chi_{\rm AB}^m(\tw)}&\\\label{eq:chiABdefsample}
&\text{with  }\chi_{\rm AB}^m(\tw)= \frac{ \sum_{ij} [ \av{ u^{AB}_i(\tw)
        u^{AB}_j(\tw)} - \av{u^{AB}_i(\tw)}\av{u^{AB}_j(\tw)} ] }{ \sum_i [\av{
        u^{AB}_i(\tw)^2} - \av{u^{AB}_i(\tw)}^2]} \ .  \end{eqnarray} 
Here, the dynamical average of $\chi_{\rm AB}^m(\tw)$ is
computed using all the $N_\mathrm{c}(N_\mathrm{c}-1)/2$ possible pairs
of A and B clones in sample $m$.     
        Both
susceptibilities, by their definition, are equal to 1 if $u_i$ and
$u_j$ are uncorrelated for all $i \neq j$, while otherwise they give
an estimate of the correlation length of particle displacements
(raised to an unknown power).

Note that while the measurement of $\D_{\rm AB}(\tw)$ and $\chi_{\rm AB}(\tw)$ requires the artificial cloning procedure, which is very difficult (if not impossible) to implement in experiments, the measurement 
of $\D(t,\tw)$ and of $\chi(t,\tw)$ is straightforwardly achievable in experiments~\cite{SD16}. In the following we will present results for both kind of observables computed always using a combination of 20-40 samples and 41-81 clones. Nevertheless, in the Appendix, we discuss in details the dependency of the statistical errors on the choice 
of $N_\mathrm{c}$ and $N_\mathrm{s}$, and the particular case (relevant for experiments), where no clone is considered.

\section{Results: hard spheres}

\subsection{Dynamics from the ergodic liquid phase}

In Fig.~\ref{fgr:dynamics}a, we report the evolution of the packing
fraction $\f(\tw)$ after a crunch from the liquid phase at initial
density $\f_{\ell}=0.565$. In this case, after a rapid growth on short
times, the packing fraction continues to evolve slowly (almost
logarithmically in $\tw$) and never reaches a stationary value.  Note
that for the largest pressure $P=323$, the compactification dynamics
starts to slow down considerably. This is due to our Monte Carlo
procedure, in which despite our careful choice of moves, the proposed
changes of volume are not accepted frequently enough at too high
pressure.  This limits the accessible range to $P\lesssim 400$.  The
values of $\f$ for the largest $\tw$ are reported in the phase diagram
of Fig.~\ref{fgr:PD}. They form a line that originates from the
equilibrium dynamical transition point $\f_{\rm d} \approx 0.594$ (the
point at which the equilibrium liquid relaxation time effectively
diverges) and ends around $\f\approx 0.66$ at infinite pressure. In
mean field, this line would correspond to the {\it threshold
  line}~\cite{CK93,MR04,Ri13,CKPUZ17} where glassy metastable states
first appear for a given pressure $P$. We would like to point out that
the determination of this line depends on the time-window studied,
although it is expected not to change too much with the original
liquid state density $\varphi_\ell$ (in the accessible time window) as long as $\varphi_\ell$ is well below $\f_{\rm d}$. As long as
$\varphi_\ell$ approaches $\f_{\rm d}$, the final point of the line,
which corresponds to the jamming point, moves to higher values of
$\f$, as discussed in Ref.~\citenum{ozawa2012jamming}. We have
checked these results by simulating a second liquid line starting from liquid configurations
at $\f_{\ell_2}=0.59$ (data can be found in the
Appendix).  As expected, the new off-equilibrium equation of state
also begins at $\f_\mathrm{d}$ but displays a slightly different slope, ending
in a slightly higher value of $\f$. Yet, the dynamics along this
second line are qualitatively very similar to those of
$\f_{\ell}=0.565$ so we will not discuss them any
further. Nevertheless, for completeness we report a figure analogous to Fig.~\ref{fgr:dynamics}
for $\f_{\ell_2}=0.59$ and $\f_{\mathrm{g}_2}=0.630$ in the Appendix.

In Fig.~\ref{fgr:dynamics}b, we report the evolution of the MSD $\D(t,\tw)$ for the same crunch from the liquid phase. 
As in the standard dynamics of glasses quenched from the liquid phase~\cite{CK93,Ca09,BB11},
upon increasing $\tw$, we observe the formation of a plateau at intermediate times $t$ in the MSD,
whose value becomes smaller with increasing $\tw$. For larger $\tw$, the plateau stabilises at its asymptotic value, but still at large $t$ the MSD is observed to depend
sensibly on $\tw$, which is the well-known phenomenon of {\it aging}~\cite{CK93,Ca09,BB11}. Older glasses (larger $\tw$) display diffusive behavior at larger times $t$.

In Fig.~\ref{fgr:dynamics}c, we report the evolution of the MSD between clones, $\D_{\rm AB}(\tw)$, for the crunch from the liquid.
We observe that $\D_{\rm AB}(\tw)$ quickly reaches large values, much larger than the individual $\D(t,\tw)$ of a single clone, indicating that
very early during
the densification process, the two clones have separated into distinct glass basins in which each of them is becoming trapped.
This is due to the fact that the initial configuration at $\f_{\ell}=0.565$ is not dynamically arrested, leaving each clone free to diffuse in a different direction
in phase space.

The results for the crunch from the liquid phase are summarised in Fig.~\ref{fgr:dynamics}d, where we report the 
values of $\D_{\rm AB}(\tw)$ and the long time limit of $\D(t,\tw)$, for several values of $\tw$, as a function of pressure $P$.
We observe that at all pressures, $\D_{\rm AB}(\tw)$ remains much larger than $\D(t,\tw)$, and the separation increases with increasing
$\tw$, corresponding to the fact that, while aging proceeds, individual clones are increasingly trapped into distinct regions of phase space.

\subsection{Dynamics from the dynamically arrested liquid phase}

In Fig.~\ref{fgr:dynamics}e, we report the evolution of the packing fraction $\f(\tw)$ after a crunch
from the dynamically arrested liquid phase, at initial density $\f_{{\rm g}_1} = 0.619$.
In this case, the dynamics is very different. 
At all pressures $P$,
 the density quickly reaches a stationary value, i.e. $\f(\tw)$ is independent of $\tw$. This long-time value is
reported in the phase diagram of Fig.~\ref{fgr:PD} as a function of pressure. This line defines the {\it equation of state} of the glass basin
selected by the initial (dynamically arrested) configuration, followed out of equilibrium as a function of the imposed 
pressure~\cite{RUYZ15}. Our results are 
consistent with a previous determination, obtained in Ref.~\citenum{BCJPSZ15}.

In Fig.~\ref{fgr:dynamics}f, we report
the dynamics of the MSD $\D(t,\tw)$ after the crunch from $\f_{{\rm g}_1}$, which is also very different than the one from the liquid.
For pressures slightly larger, but close to the initial equilibrium value (e.g. $P=46$), after a short initial transient,
no aging is observed. The MSD becomes stationary, i.e. $\D(t,\tw)$ is independent of $\tw$, but it shows no diffusive behavior at large times~$t$. Instead, 
the plateau extends at infinite times, indicating that the system has reached 
equilibrium within a restricted portion of phase space that defines a {\it glass basin}~\cite{PZ10,CKPUZ17}. The same behavior is observed for all pressures
$P \lesssim 97$. For pressures $P\approx 97$ and above, a different behavior is observed: the plateau does not extend to infinite times, but at large values
of $t$ the MSD departs from the plateau to reach higher values, in a $\tw$-dependent manner. However, a fully diffusive regime is never observed, and the MSD seems
to saturate at a higher plateau. These results are consistent with the expectation based on mean field theory in presence of a Gardner transition~\cite{CKPUZ17}, and with the numerical
results of Refs.~\citenum{CJPRSZ15,BCJPSZ15}.

In Fig.~\ref{fgr:dynamics}g, we report
the dynamics of the MSD between different clones, $\D_{\rm AB}(\tw)$. We find that, differently from the crunch from the liquid, in this case 
$\D_{\rm AB}(\tw)$ overshoots, as an effect of the compression, and then saturates to a small value, that is always of the same order of magnitude as $\D(t,\tw)$. In particular, for the lower pressures $P \lesssim 97$,
the long $\tw$ limit of $\D_{\rm AB}(\tw)$ coincides with the long $t$ limit of $\D(t,\tw)$ (which is independent of $\tw$ for large enough $\tw$),
indicating that the glass basins is sampled ergodically~\cite{FP95,RUYZ15,CJPRSZ15,BCJPSZ15,CKPUZ17}.
Around $P\approx97$, the long time limits of $\D_{\rm AB}(\tw)$ and $\D(t,\tw)$ separate, while remaining of the same order of magnitude,
indicating that the two clones are falling into distinct sub-regions of the same glass basin~\cite{CJPRSZ15,BCJPSZ15,SD16,SBZ17,SRPZ18}.

To highlight this separation process, in Fig.~\ref{fgr:dynamics}h
we report the 
values of $\D_{\rm AB}(\tw)$ and the long time limit of $\D(t,\tw)$, for several values of $\tw$, as a function of pressure $P$.
The two values coincide at all $\tw$s for $P \lesssim 97$ (with the exception of the liquid point where some remanent diffusion is observed at the longest times), while they separate above this value, their separation increasing upon increasing $\tw$.
This indicates that a sharp dynamical crossover is taking place around $P\approx 97$, which defines the (average)
Gardner crossover point for the glass
basins selected at initial
density $\f_{{\rm g}_1}$.
The results for the other initial density $\f_{{\rm g}_2}$, which can be found in the Appendix, are qualititatively similar, with the only difference that the Gardner crossover
is shifted to higher pressures~\cite{BCJPSZ15}.

In summary, the analysis of the dynamical evolution of the density and of the MSD after a sudden crunch allows one to clearly distinguish between
{\it (i)} the standard aging from the liquid phase at $\f_{\rm \ell}$, 
which is observed at all pressures and corresponds to each clone sampling independently the emergence
of distinct metastable ``threshold'' states, and {\it (ii)} the aging from a dynamically arrested liquid configuration at $\f_{{\rm g}_1}$ or $\f_{{\rm g}_2}$,
which is only observed at pressures larger than a well-defined crossover point, and corresponds to the same glass basin breaking into a set of distinct sub-regions,
explored by distinct clones.

\begin{figure}[t]
	\centering
	\includegraphics[height=6cm]{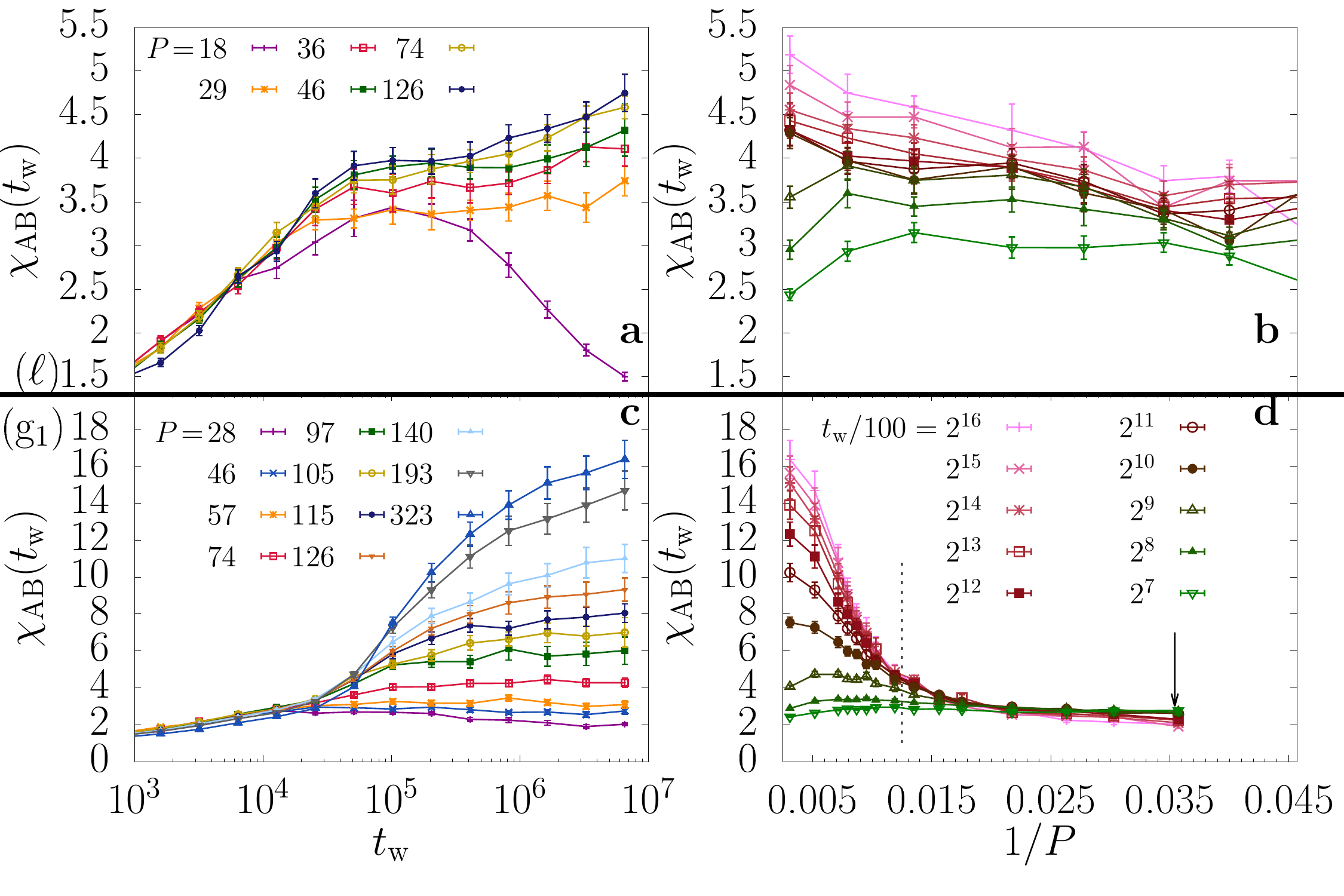}
	\caption{{\bf Clone (spin-glass) susceptibility.} We show the values
		of the susceptibility defined in Eq.\eqref{eq:chiABdef} for the
		ergodic liquid ($\varphi_\ell=0.565$) as function of $\tw$ in (a) and for different $\tw$ as
		function of $1/P$ in (b). In the bottom panels (c,d) we show the same
		curves for the dynamically arrested liquid ($\varphi_\mathrm{g1}=0.619$). 
		While $\chi_{\rm AB}$ is essentially insensitive to all
		$\tw$ and $P$ in the liquid, in the glass this behavior is only
		observed at low pressures, and a strong growth of the susceptibility
		is observed for $P\gtrsim 97$. As a guide to the eye, we have
		included the value of the position of the Gardner crossover for this $\varphi_\mathrm{g}$
		estimated in Ref.~\citenum{BCJPSZ15} as a vertical dashed line, and an arrow to indicate the initial pressure. Most of the susceptibilities were computed using $N_\mathrm{s}=20$ samples and $N_\mathrm{c}=41$ clones, with the exception of the curves for $\varphi_{\mathrm{g}_1}$ for $P\ge 115$, where  $N_\mathrm{s}=40$ samples and $N_\mathrm{c}=81$ clones were used.} 
	\label{fgr:suscAB}
\end{figure}

\begin{figure}[t]
	\centering
	\includegraphics[height=4cm]{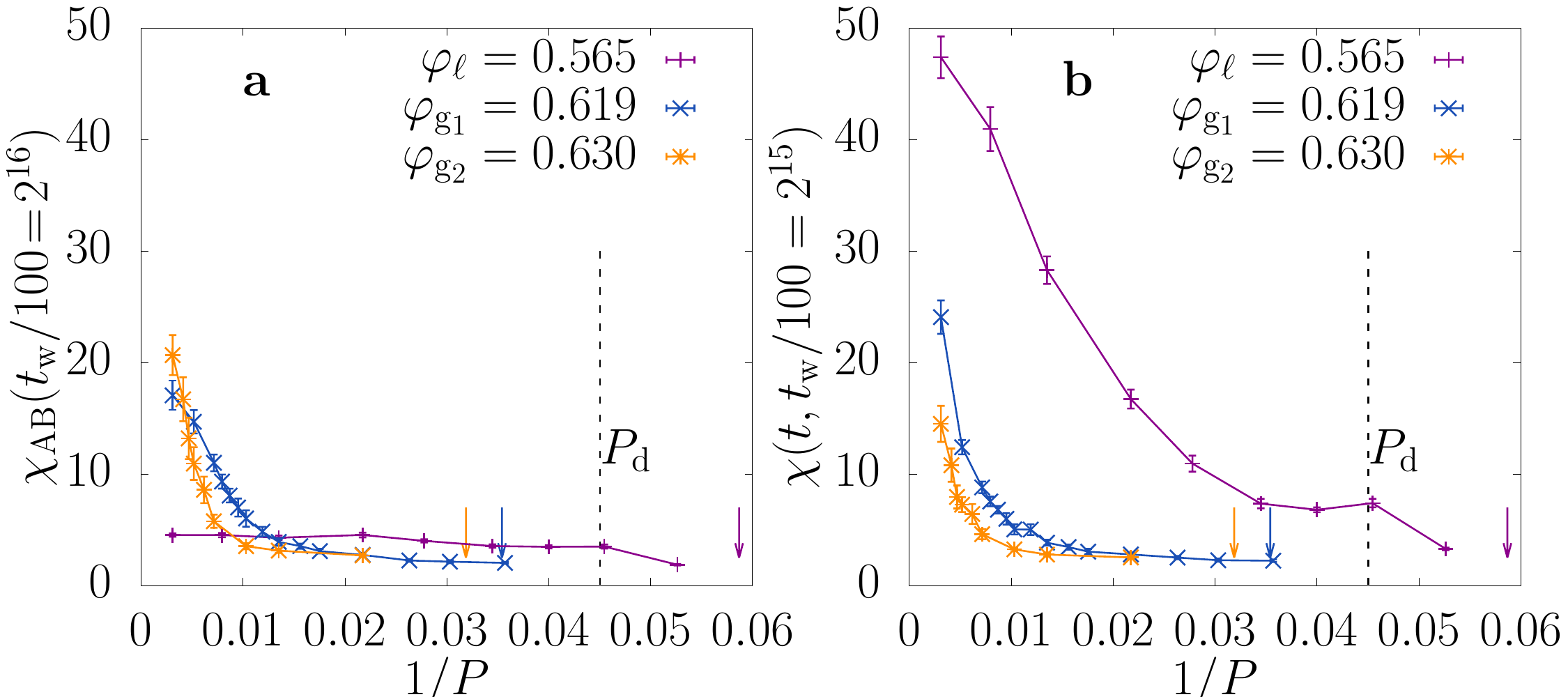}
	\caption{{\bf Susceptibilities along the compression.} We show $\chi_{\rm AB}(\tw)$ (a) and $\chi(t,\tw)$ (b) for the largest accessible times,
		as functions of $1/P$, for the three compression lines studied. We include the position of the dynamical transition $P_\mathrm{d}$ as a vertical dashed line.
	}
	\label{fgr:suscAB2}
\end{figure}

\begin{figure}[t]
	\centering
	\includegraphics[height=6cm]{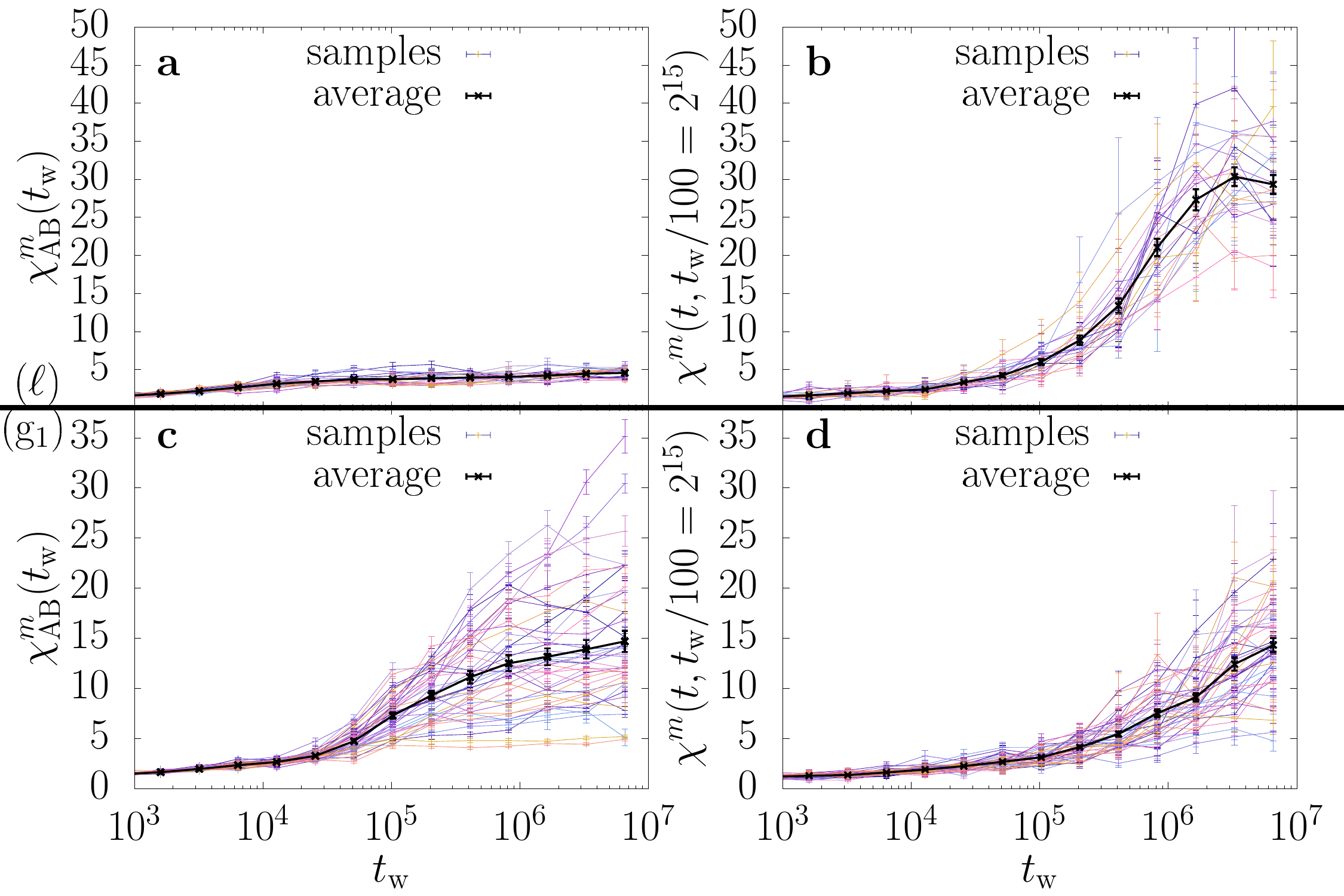}
	\caption{{\bf Sample-to-sample fluctuations.} We show the individual sample estimations of (a,c) $\chi^m_{\rm AB}(\tw)$, defined in Eq.~\eqref{eq:chiABdefsample}, and 
	(b,d) $\chi^m(t,\tw)$, defined in Eq.~\eqref{eq:chidyndefsample}, for $\varphi_\ell$ at $P=74$  and $\varphi_{{\rm g}_1}$ at $P=193$ obtained by averaging over 41 and 81 clones respectively. The sample averages are also shown as thick black line-points, which are also reported in Fig.~\ref{fgr:suscAB}. 
	}
	\label{fgr:susc-samples}
\end{figure}

\begin{figure}[t]
\centering
  \includegraphics[height=6cm]{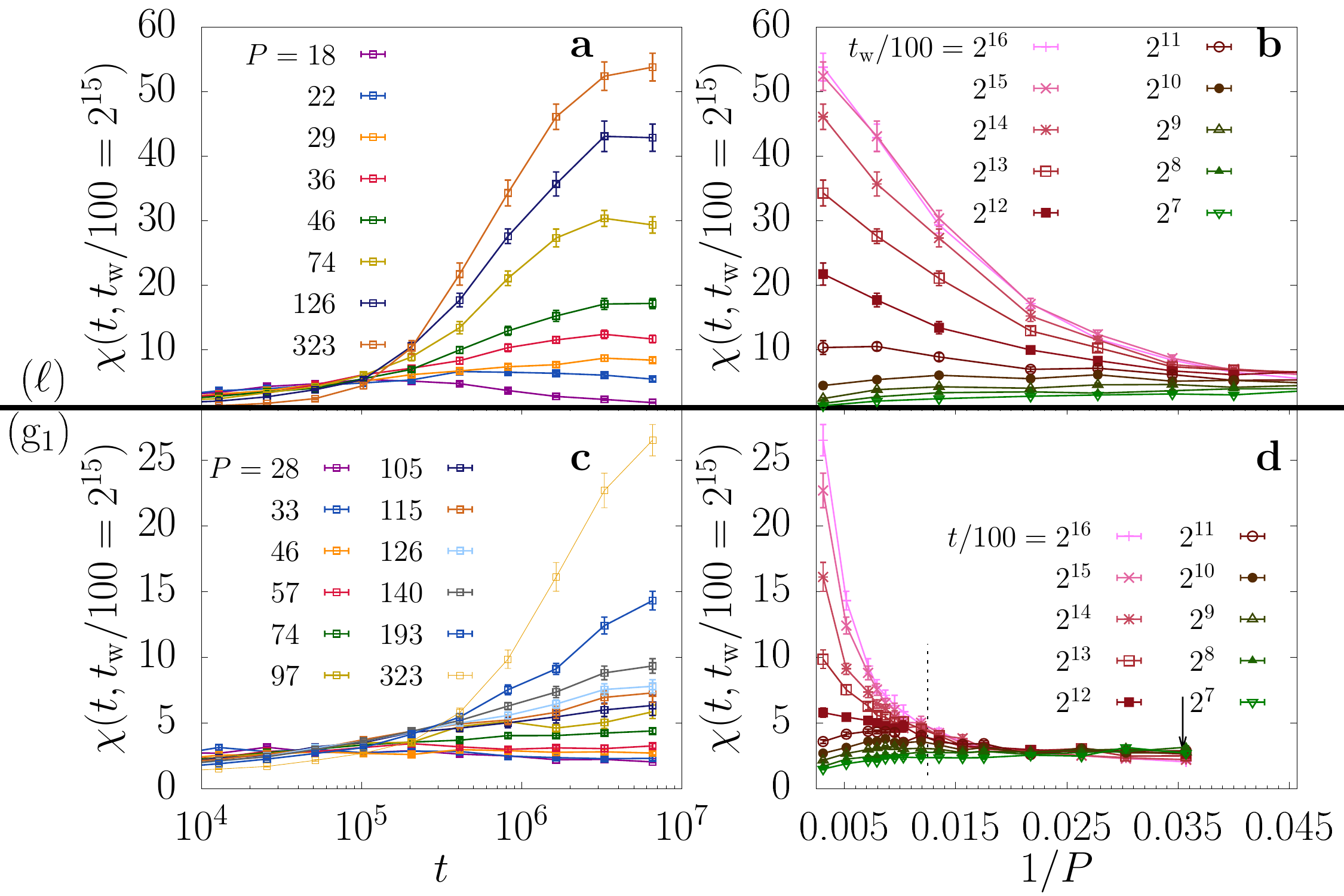}
  \caption{{\bf Dynamical susceptibility.} We show the values of the
    dynamical susceptibility defined in Eq.\eqref{eq:chiABdef}, for the largest available $\tw$, as
    function of $t$ in (a,c) and for different $t$ as function of
    $1/P$ in (b,d). Contrary to Fig.~\ref{fgr:suscAB}, a strong
    dependence of $\chi$ with $P$ and $\tw$ is observed in the crunch from the ergodic
    liquid (a,b), which corresponds to the well-known dynamical
    heterogeneities. The curves for the dynamically arrested liquid
    ($\varphi_\mathrm{g1}=0.619$) are shown in (c,d) displaying the
    same qualitative behavior than~$\chi_{\rm AB}$. These quantities are computed using  $N_\mathrm{s}=20$ samples and $N_{\mathrm{c}}=41$ clones, 
    with the exception of the curves for $\varphi_{\mathrm{g}_1}$ and $P\ge 105$, where  $N_{\mathrm{s}}=40$ samples were used.      }
  \label{fgr:susc-dyn}
\end{figure}

\subsection{Clone (spin-glass) susceptibility}

The structure of sub-regions observed within the glass basin might be due to localised defects, i.e. small groups of particles that
can be in two distinct local energy minima, leading to small barriers~\cite{Go69,heuer2008exploring,SBZ17,SRPZ18}, 
or to extended defects, leading to collective rearrangements and
higher barriers, as predicted by mean field in spin-glasses~\cite{MPV87} and in particle systems~\cite{CKPUZ14NatComm,CKPUZ17}.
A good way to discriminate between these two scenarios is to investigate the susceptibility $\chi_{\rm AB}(\tw)$ defined in Eq.~\eqref{eq:chiABdef},
associated to the displacement between distinct clones, also
known as spin-glass susceptibility in spin-glass physics.
A small value of $\chi_{\rm AB}(\tw)$ indicates small spatial correlations, suggesting localised defects~\cite{SBZ17,SRPZ18}, while 
a large $\chi_{\rm AB}(\tw)$ indicates large-scale spatially correlated, collective excitations~\cite{BCJPSZ15}.

In Fig.~\ref{fgr:suscAB}a we report
$\c_{\rm AB}(\tw)$ as a function of $\tw$ for several pressures,
for a crunch from the ergodic liquid at initial density $\f_{\ell}=0.565$. In Fig.~\ref{fgr:suscAB}b, the same data are plotted as a function of
$1/P$ for several values of $\tw$. For the lowest pressure $P=19$, $\chi_{\rm AB}(\tw)$ grows but then reaches a maximum and decreases,
indicating full decorrelation of the clones at long times.
This behavior is reminiscent of the well-studied dynamical heterogeneities that characterise the liquid phase
upon approaching the dynamical transition~\cite{BB11,BBBCS11}. For larger pressures, the susceptibility is always increasing with $\tw$, but
the growth is modest (approximately a factor of 2) and roughly
independent of pressure. 
In Fig.~\ref{fgr:suscAB}b it is clearly seen that the aging is qualitatively similar at all pressures, and no crossover is detected.
This is consistent with Fig.~\ref{fgr:dynamics}b and
with the standard properties of aging after a quench from the liquid phase, in which distinct clones are expected to explore distinct (hence uncorrelated) threshold states~\cite{CK93}. Note that the 
existence of a Gardner-like crossover in this regime
has been proposed in Ref.~\citenum{Ri13}, but the crossover is expected to be very weak and difficult to detect in numerical simulations.
We leave a more detailed investigation of this crossover for future work.

A very different behavior is observed when the crunch is from the dynamically arrested phase, at $\f_{{\rm g}_1} = 0.619$, as reported in
Fig.~\ref{fgr:suscAB}c. In this case, at low pressure $\chi_{\rm AB}(\tw)$ is roughly constant, while at high pressure we observe a sharp and
continuous increase of $\chi_{\rm AB}(\tw)$ upon increasing $\tw$, by a factor of $\approx 10$ for the largest pressures, over the accessible time window. The growth
is initially faster, and then crosses over to a slower regime, roughly logarithmic in $\tw$.
The same data are shown in Fig.~\ref{fgr:suscAB}d as a function of $1/P$ for several $\tw$, which makes clear that there is a sharp crossover
around $P\approx 97$, as observed in the MSD. For $P\lesssim 97$, no growth of $\chi_{\rm AB}(\tw)$ is observed, while a sharp growth is observed for
$P \gtrsim 97$. While we do not observe a sharp divergence of the susceptibility, the crossover point is roughly consistent with the values 
reported in Ref.~\citenum{BCJPSZ15}, where the crossover was estimated by a different method.

To summarize these results, in Fig.~\ref{fgr:suscAB2}a we report $\chi_{\rm AB}(\tw)$ for the largest $\tw$, as a function of inverse pressure $1/P$,
for the crunches from the three initial states at $\f_{\rm \ell}$, $\f_{{\rm g}_1}$, and $\f_{{\rm g}_2}$. This plot clearly shows that no correlation between clones
is observed in the crunch from the ergodic liquid, while a large correlation builds up in the crunches from the dynamically arrested phase, below the Gardner 
crossover. Furthermore, the Gardner crossover is found to shift at higher pressures for denser initial states, consistently with 
the theory~\cite{KPUZ13,RUYZ15,CKPUZ17}
and with earlier numerical 
results on the same system~\cite{BCJPSZ15} illustrated in Fig.~\ref{fgr:PD}.

Before concluding this Section, we would like to stress that despite the relatively smooth behavior of $\chi_{\rm AB}(\tw)$, we find very strong sample-to-sample 
fluctuations in the estimations of this quantity on each individual sample, that is, in the $\chi^m_{\rm AB}$ defined in Eq.~\eqref{eq:chiABdefsample}. 
Furthermore, we observe diverse qualitative evolutions of  $\chi^m_{\rm AB}$ with $\tw$, which can: (i)~grow fast with time, 
(ii)~grow at short times and decrease afterwards or (iii)~remain essentially constant. 
We show the evolution of these individual sample susceptibilities in  Fig.~\ref{fgr:susc-samples}(a,c), but we leave a more systematic investigation of
sample-to-sample fluctuations for future work.

\subsection{Dynamical susceptibility}

The clone susceptibility $\chi_{\rm AB}(\tw)$ is easily studied in numerical simulations, but 
it is very hard to measure in experiments, because one cannot prepare two clones of the system
starting in the same initial configurations with different velocities. A solution that was adopted in
the granular experiment of Seguin and Dauchot~\cite{SD16} was to perform compression/decompression 
cycles.
Here, we discuss another possible solution, which consists in investigating the dynamical susceptibility
$\chi(t,\tw)$ for a single clone of the system. This observables is straightforward to measure in experiments,
provided one is able to repeat the experiment a sufficient number of times to obtain a proper averaging
over clones or samples.

In Fig.~\ref{fgr:susc-dyn} we report the dynamical susceptibility, for a fixed value of $\tw$ (the largest one we can access),
as a function of time $t$. 
In the case of a crunch from the ergodic liquid (Fig.~\ref{fgr:susc-dyn}a), the system performs standard aging, and the dynamical susceptibility
is different from the clone one: it displays a maximum as a function of $t$, and then decreases, at least for the lowest pressures. For the higher pressures,
we believe that the time scale at which $\chi(t,\tw)$ decreases has fallen outside the accessible time window. 
Also, the susceptibility
reaches larger values. This is associated to the well know dynamical heterogeneities that
develop in this regime~\cite{BB11,BBBCS11}. When plotted as a function of inverse pressure $1/P$ (Fig.~\ref{fgr:susc-dyn}b), 
the susceptibility is found to be time-dependent
 at all pressures, and no sharp crossover is visible, confirming that in this case the aging is qualitatively similar at all pressures.

Instead, in the case of a crunch from the dynamically arrested phase (Fig.~\ref{fgr:susc-dyn}b), the behavior of $\c(t,\tw)$
is qualitatively very similar to the one of the clone susceptibility reported in Fig.~\ref{fgr:suscAB}.
When plotted as a function of inverse pressure $1/P$ (Fig.~\ref{fgr:susc-dyn}d), the susceptibility is found to increase with $t$ at pressures larger
than the Gardner crossover, and to remain small and independent of $t$ at pressure below it.
We thus conclude that in this case a dynamical susceptibility measurement, for an appropriate value of $\tw$, provides similar information to
a measurement of the clone susceptibility. A direct comparison of the two susceptibilities is given in Fig.~\ref{fgr:suscAB2}.
We also show the sample-to-sample fluctuations of this observable in Fig.~\ref{fgr:susc-samples}(b,d).

\begin{figure}[t]
\centering
  \includegraphics[height=3.5cm]{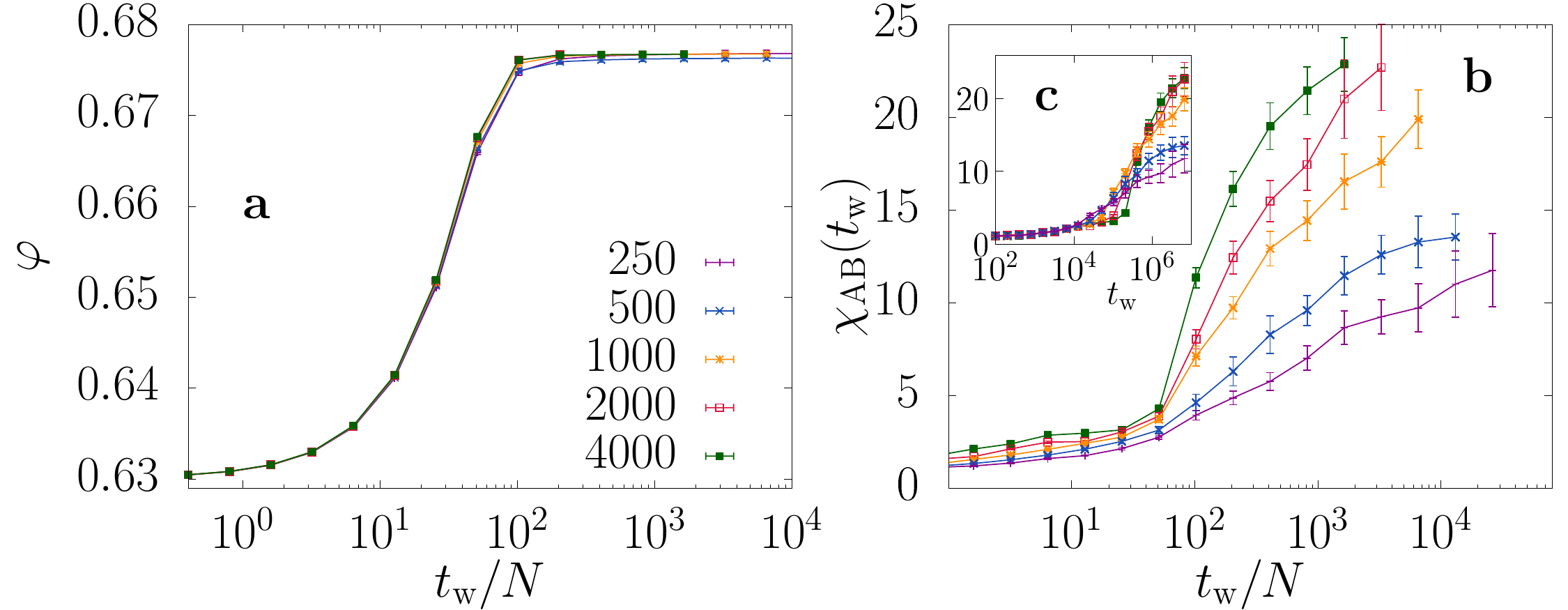}
  \caption{{\bf Finite size effects.} For the case $\f_{\mathrm{g}_2}=0.630$, we show in (a) the evolution of packing fraction $\varphi$ for different system sizes $N$ as a
  function of the scaled variable $\tw/N$. The curves collapse, 
  showing that the compression rate is $N$ dependent in our set-up. In figure (b) we show $\chi_{\rm AB}$ for different $N$ as function of the same variable, showing that the curves collapse during the compression (short times) but not at longer times, where the density stabilises to its final value.
  In this regime, larger $N$ have slower compression rates, hence older effective age, corresponding to a higher $\chi_{\rm AB}$. This effect
  might be mixed with genuine finite size effects.
  In the inset (c) we show the same curves but plotted versus the non-scaled $\tw$, showing that the higher system sizes roughly collapse to the same curve. In all cases, $N_\mathrm{s}=20$ and $N_\mathrm{c}=41$ were used. }
  \label{fgr:FS}
\end{figure}

\subsection{Finite size effects}\label{sec:FS}

In Fig.~\ref{fgr:FS} we discuss the robustness of our results against
finite size effects by varying the number of particles $N$.  As
discussed in Section~\ref{sec:simu}, the constant-pressure Monte Carlo
code uses moves of $\d V\propto 1/N$ in order to keep a constant
acceptance rate.  As a result, the compression rate scales as
$1/\d V \propto N$ as we show in Fig.~\ref{fgr:FS}a where the
densification process is found to be independent of system size, once
the time is rescaled by $N$.  We show $\chi_{\rm AB}(\tw)$ as function
of the same scaling variable in Fig.~\ref{fgr:FS}b, showing that all
curves collapse at short times, during the compression process.
At longer times, once $\f(\tw)$ has stabilised, larger system display
a faster growth of the susceptibility, which is unexpected. Indeed, as
long as the correlation length $\xi(\tw)$ remains much lower than the
linear size of the simulation box $L$, the growing of the amorphous
domains should be independent of $N$. 
It might be that the linear size is not large enough to eliminate completely the finite size effects:
indeed, in spin-glasses it has been estimated that one
needs to simulate systems with $L>7\xi(\tw)$ in order to avoid these
effects~\cite{janus2008nonequilibrium}.
This observation might also be
related to our simulation set-up: larger systems have slower compression rate, hence larger effective
age, corresponding to larger $\chi_{\rm AB}(\tw)$. 
Indeed, while the compression is
$N$-dependent, the exploration of the cages is not, which makes hard
to compare the different system sizes at the same physical time. In
fact, if we plot the same curves as a function of $\tw$ instead of
$\tw/N$ (Fig.~\ref{fgr:FS}c), we can see that the curves for
the two largest $N$ collapse. 
Based on this discussion, we believe that the use of the value of $N=1000$ is enough 
for the purpose of our investigation. 

\section{Results: spin glass}

\begin{figure*}[t]
\centering
  \includegraphics[height=7cm]{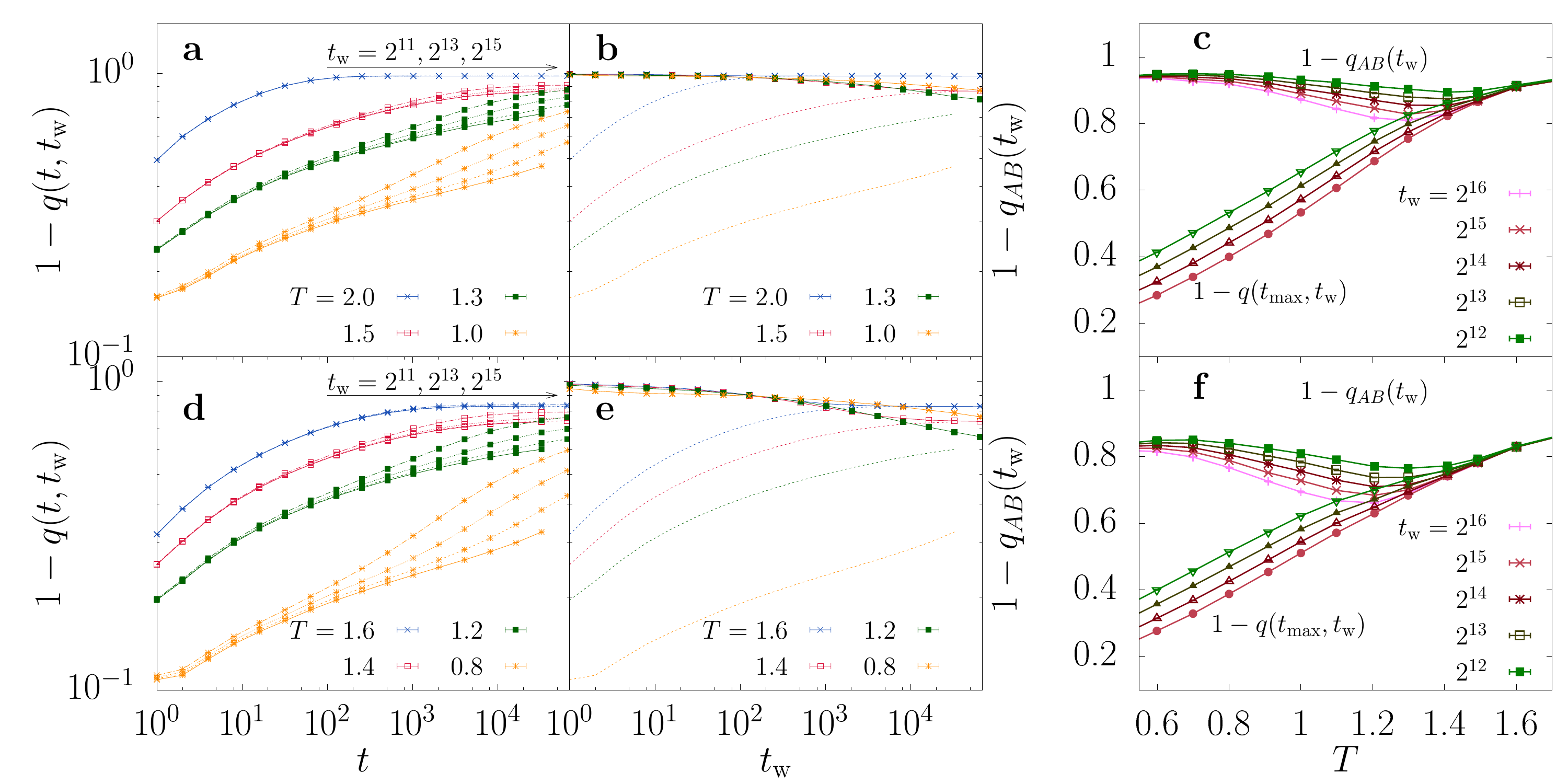}
  \caption{{\bf Overlap dynamics after a sudden quench for the spin
      glass in a field.} We show the dynamics of $1-q(t,\tw)$ and
    $1-q_{\rm AB}(\tw)$ for two different values of the external
    field, $H=0.1$ (top panels) and $H=0.2$ (bottom panels). In
    figures (a,d) we plot $1-q(t,\tw)$ as function of $t$ for
    different values of $\tw$ after the instantaneous quench in
    temperature. In figure (a) aging is visible for
    $T\lesssim 1.6$ while in (d) it appears at
    $T\lesssim 1.4$. In (b,e) we show 
    $1-q(t,\tw)$ as function of $t$ for the highest available $\tw$, together with
    $1-q_{\rm AB}(\tw)$ as function of $\tw$. One can observe that
    both curves converge to the same asymptote for temperatures above the
    onset of aging and differ by much below it. In panels (c,f) we
    show the value of $1-q(t,\tw)$ corresponding to the largest time in (b,e), together with $1-q_{\rm AB}(\tw)$ as a function of $T$ for several $\tw$, which
   shows the separation at the onset of aging.  }
  \label{fgr:dynSG}
\end{figure*}

\begin{figure}[t]
\centering
  \includegraphics[height=6cm]{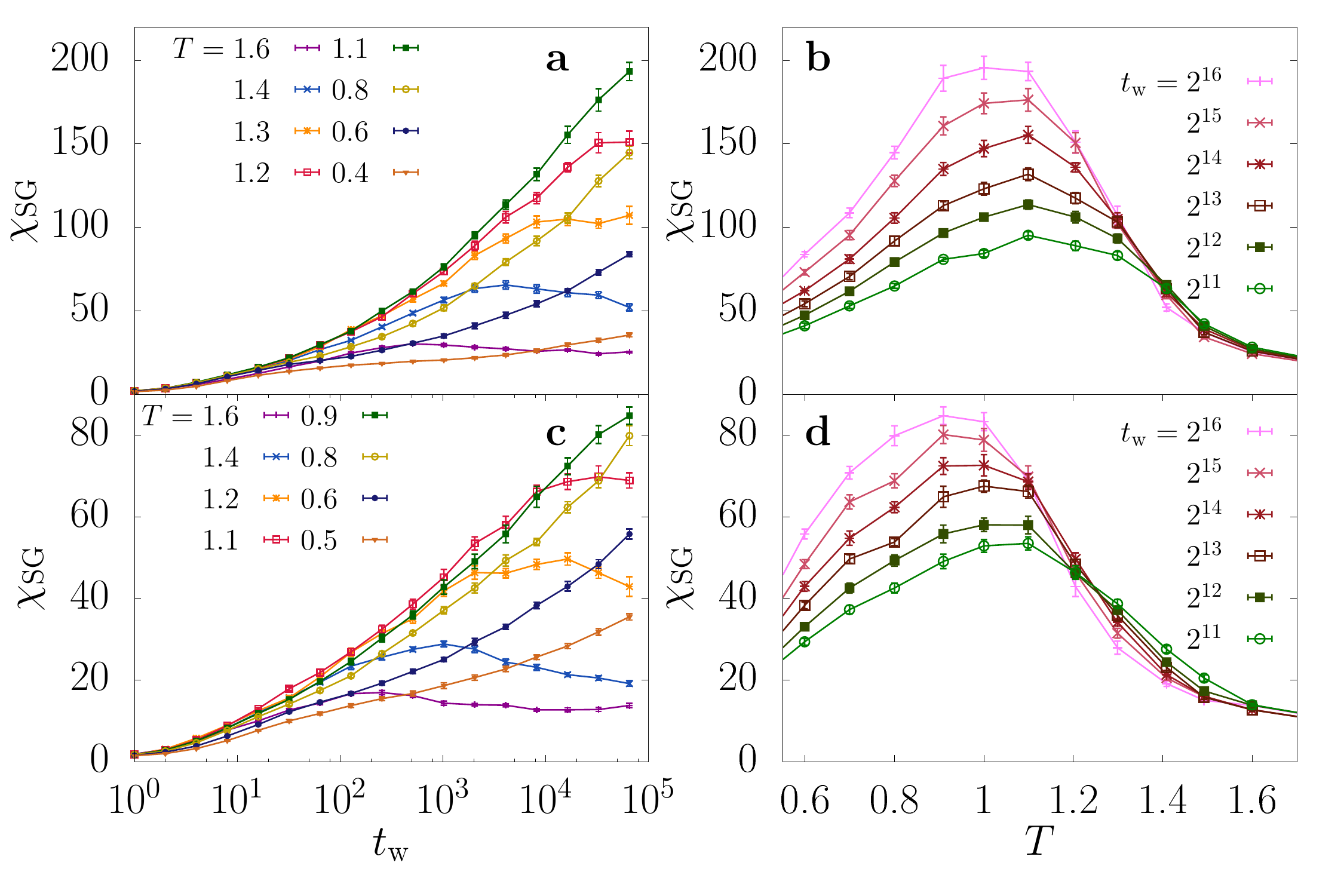}
  \caption{{\bf Susceptibility for the spin glass in a field.} We show results for the spin glass susceptibility in a field defined in Eq.~\eqref{eq:chiSGdef} for two different values of the field $H=0.1$ (top panels) and $H=0.2$ (bottom panels). In panels (a,c) we show the evolution with $\tw$ for different temperatures and in (b,d) the same data but plotted versus $T$ for several $\tw$.
  }
  \label{fgr:suscSG}
\end{figure}

\subsection{Details of the simulation}

To obtain a direct comparison of hard spheres and a spin-glass in a field, we performed 
Monte Carlo simulations of an Edwards-Anderson spin-glass model in an external magnetic field~\cite{MPV87}, using the
Metropolis algorithm.  
Our $3d$ system is defined on a cubic lattice with linear size $L=24$ and periodic boundary conditions.
The Hamiltonian is
\beq
\HH = - \sum_{\langle i,j \rangle} J_{ij} s_i s_j  - H \sum_i s_i \ ,
\eeq
where $s_i = \pm 1$ are spin variables located on the vertices of the lattice,
the first sum is over nearest neighbors on the lattice, and
the $J_{ij}$ are i.i.d. quenched random variables that take the value $\pm 1$ with equal probability. 
We studied the system with fields $H=0.1$ and $H=0.2$.
For each value of temperature and field, we simulate $N_{\rm s}=1280$ samples (independent realisation of the disorder $J_{ij}$)
and $N_{\rm c}=4$ clones. Each clone is prepared at $\tw=0$ at infinite temperature, i.e. in a random spin configuration, and instantaneously
quenched at the final temperature $T$, where its dynamical evolution is followed.

\subsection{Mapping between spin-glasses and structural glasses}

In Fig.~\ref{fgr:dynSG} we report the overlap dynamics of the spin-glass, where the overlap $q(t,\tw) = (1/N)\sum_i \langle s_i(t+\tw)s_i(\tw)\rangle$ is
equal to 1 for identical configurations $\{s_i(\tw)\}$ and $\{s_i(\tw+t)\}$, and smaller than 1 otherwise.
The quantity $1-q(t,\tw)$ is then equivalent of the mean-square displacement for particle systems, as 
both quantities measure the distance in phase space between configurations, and both
 increase for increasingly different configurations.
The same reasoning can be applied to the clone overlap $q_{\rm AB}(\tw) = (1/N)\sum_i \langle s^{\rm A}_i(\tw)s^{\rm B}_i(\tw)\rangle$.

For both values of the external field $H$, we observe that at high temperature the clone overlap converges to a value independent of $\tw$, while
the dynamical overlap becomes $\tw$ independent and its long time limit $q(t\to\io, \tw\to\io)$ coincides with the one of $q_{\rm AB}(\tw\to\io)$
(Fig.~\ref{fgr:dynSG}a,d).
This implies that phase space is sampled ergodically in the high temperature phase.
Upon lowering the temperature, for both values of $H$, we observe that a second plateau emerges in $q(t,\tw)$, and aging appears at large values
of $t$ (Fig.~\ref{fgr:dynSG}a,d). Correspondingly, the long-time limits of $q(t,\tw)$ and $q_{\rm AB}(\tw)$ separate (Fig.~\ref{fgr:dynSG}b,e).
The separation becomes more evident when the long time limits are plotted as a function of temperature (Fig.~\ref{fgr:dynSG}c,f).
The dynamical crossover temperature is lower for larger values of magnetic field.

Based on these results,
let us discuss briefly the mapping between structural glasses and spin-glasses in an external field, 
referring to Refs.~\citenum{FM13,BU15} for more details.
In spin-glasses, a realisation of the quenched disorder $J_{ij}$ defines a sample. Each sample has a single ergodic (paramagnetic) free energy 
basin at high temperature, and undergoes a crossover to a non-ergodic fractured (spin-glass) basin at low temperatures.
In structural glasses, the role of quenched disorder is played by the initial condition, provided one is in the dynamically arrested phase. In that
phase, each initial configuration belongs to a single glass basin, that can be identified with the paramagnetic basin of the spin-glass. Upon increasing
pressure, the glass basin fractures into a non-ergodic basin, akin to the spin-glass phase.
The similarity of the spin-glass results in Fig.~\ref{fgr:dynSG} with the structural glass ones in Fig.~\ref{fgr:dynamics} (lower line, corresponding to 
a crunch from the dynamically arrested liquid) confirm this picture.

Note that this reasoning does not hold for the structural glass if the initial configuration is in the ergodic liquid phase. In that case, the initial configuration
does not belong to a glass free energy basin: it is free to explore all the liquid phase space, and upon compression the physics is very different, as metastable
states form abruptly and trap the dynamics giving rise to a standard discontinuous glass transition. 
The observed phenomenology (Fig.~\ref{fgr:dynamics}, upper line) is, indeed, different.

\subsection{Spin-glass susceptibility}

In Fig.~\ref{fgr:suscSG}a,c we report the spin-glass susceptibility 
\beq\label{eq:chiSGdef}
\chi_{\rm SG}(\tw) = 
\overline{\left[  \sum_{ij} [ \av{ u_i(\tw) u_j(\tw)} - \av{u_i(\tw)}\av{u_j(\tw)} ] \right]} \ ,
\eeq
with $u_i(\tw) = s_i^{\rm A}(\tw) s_i^{\rm B}(\tw)$,
which is the equivalent of the
clone susceptibility for the hard sphere case.
For both values of the field $H$, we observe that at high temperature the susceptibility is roughly constant. Upon lowering the temperature,
the susceptibility starts increasing with $\tw$, and it saturates to a larger and larger value.
At some point, the saturation time exceeds the accessible time of the simulation, and the susceptibility constantly increases in a roughly
logarithmic manner. For even lower temperature, we observe that the increase remains roughly logarithmic, but with a smaller and
smaller prefactor, indicating that the aging slows down. The latter phenomenon is not observed in our hard sphere samples. We presume
that this is due to the fact that we cannot reach sufficiently large pressures.

The same data can be plotted as a function of $T$ for various $\tw$ (Fig.~\ref{fgr:suscSG}b,d).
For small $\tw$, the susceptibility 
displays a maximum around the point where aging begins (the equivalent of the Gardner crossover),
once again due to the fact that the aging slows down at very low temperatures. 
Upon increasing $\tw$, the maximum grows and shifts progressively to lower temperatures.
Note that this maximum is only observed in the hard sphere case
at short $\tw$ (see e.g. the data for $\tw/100= 2^9$ in Fig.~\ref{fgr:suscAB}d), presumably, as already mentioned, because of the limited range of
pressures: upon increasing $\tw$ the maximum shifts to higher pressures and falls out of the accessible range.

\section{Conclusions}

In this paper, we studied the dynamics of a polydisperse
$3d$ hard sphere glass. The glass was prepared either in the ergodic liquid phase, at densities below
the dynamical (or MCT) transition,
or in the dynamically arrested phase above it. Then, the system was instantaneously crunched to a higher pressure, and the subsequent
aging dynamics was investigated.

The crunch dynamics from the ergodic liquid phase is in agreement with previous studies performed in this regime.
Different realisations of the dynamics (clones) evolve towards decorrelated {\it threshold} states, that progressively trap the dynamics.
The distance between clones $\D_{\rm AB}(\tw)$ is large, while the mean square displacement $\D(t,\tw)$ of a single clone displays
aging at all pressures. The clone susceptibility $\chi_{\rm AB}(\tw)$ is featureless, while the dynamical susceptibility $\chi(t,\tw)$ grows
at all pressures indicating the presence of dynamical heterogeneities, which, in mean field, is due to the criticality 
of the threshold states~\cite{CK93}.

The crunch dynamics from the dynamically arrested phase is instead very different. Here, initial configurations are already trapped
in a glass basin before the crunch. If the crunch is to a pressure moderately higher than the initial one, $P \lesssim P_{\rm G}$,
then one observes a fast relaxation dynamics to a new state in which the glass basin is ergodically sampled at the higher pressure.
In this case, no aging is observed and the susceptibilities remain small. Furthermore, the long time limits of $\D_{\rm AB}(\tw\to\io)$ and
$\D(t\to\io,\tw\to\io)$ coincide, and the same happens for the susceptibilities.
By crunching at higher pressures $P \gtrsim P_{\rm G}$, we observe instead that the glass basin becomes fractured in sub-basins, leading
to the emergence of long time scales, and the associated aging. At the same time, the susceptibilities grow, indicating the cooperative nature
of the excitations that lead to crossing barriers between sub-states in a basin.

We identify the pressure $P_{\rm G}$ as a Gardner crossover, reminiscent of the Gardner phase transition predicted by 
mean field theory~\cite{KPUZ13, CKPUZ17}.
However, we note that in our $3d$ simulations the growth of the susceptibility $\chi_{\rm AB}(\tw)$ with $\tw$ 
is very slow (logarithmic in time), which hinders the possibility
of observing large values of $\chi_{\rm AB}(\tw)$ and thus deciding whether $P_{\rm G}$ is a real phase transition or just a 
sharp dynamical crossover.

The situation is very similar to the case of spin-glasses in an external magnetic field, where mean field predicts a sharp second-order
phase transition~\cite{MPV87}, 
while numerical simulations observe a slowly growing spin-glass susceptibility~\cite{janus2014dynamical} and are unable to decide on the existence
of a transition~\cite{janus2014three}. 
When plotting $\chi_{\rm SG}(\tw)$ as a function of $T$, one observes a maximum that slowly grows with increasing $\tw$
(suggesting a phase transition), but also slowly shifts to lower temperatures (suggesting that the critical point might be only at $T=0$).
Unfortunately, both processes happen very slowly (logarithmically) with increasing $\tw$, so that it is not possible to extract reliably the 
asymptotic behavior. While on the observational time scales, it is clear that some cooperative phenomenon is taking place, deciding on
whether this phenomenon is of thermodynamical or dynamical nature is very difficult.
Our data strongly suggest that the situation is very similar in the $3d$ hard sphere glasses, supporting the idea that the two systems could be in the same
universality class, as suggested by theoretical arguments~\cite{FM13,BU15,BU16,charbonneau2017nontrivial}. 
Unfortunately, due to the limited pressure range of the hard sphere simulation,
the maximum in $\chi_{\rm AB}(\tw)$ is less evident than in the spin-glass case.

Finally, we note that in the crunches from the dynamically arrested liquid phase, we find strong indications that the system is not able to leave the original
glass basin in which it is trapped in equilibrium: the particle displacements always remain very small (less than a tenth of the average particle
diameter, see Fig.~\ref{fgr:dynamics}f). Thus, no activated events in which the system jumps among distinct glass basins
are observed in our simulations, which prevents us from testing the ideas developed in Ref.~\citenum{LW18} about 
RFOT-like nucleation and the associated buckling instabilities in hard sphere glasses.

We conclude that $3d$ colloidal and granular glasses display
spin-glass like aging beyond the Gardner crossover, located deeply in
the off-equilibrium glass phase.  The aging is associated to collective excitations
that are responsible for many anomalies of the solid phase.
Unfortunately, as in the spin-glass, equilibration times grow very
fast, and a systematic study of the crossover as a function of system
size and $\tw$ is impossible with current computers.  Yet, an
interesting direction for future work would be to perform more
systematic studies of the sample-to-sample fluctuations, that are
known to be highly non-trivial in spin-glasses.  Another interesting
future direction would be to study the nature of the Gardner
transition in equilibrium~\cite{BU15} as in the Edwards-Anderson
model~\cite{janus2014three,jorg:08b}, using the replica
exchange algorithm (parallel tempering) for small systems~\cite{swendsen1986replica,marinari1992simulated}
(with temperature replaced by pressure).
This might shed additional light on the nature of the crossover.

\section*{Acknowledgements}
We wish to thank Ludovic Berthier, Patrick Charbonneau, Olivier Dauchot, Yuliang Jin, Qinyi Liao, Mike Moore, Misaki Ozawa, Giorgio Parisi,
Camille Scalliet, Pierfrancesco Urbani, Peter Wolynes, and Sho Yaida for many useful discussions. 
We thank Yuliang Jin for providing us the initial equilibrated configurations from Ref.~\citenum{JY17}. We thank
the {\em Janus collaboration}, and in particular Marco Baity-Jesi, Luis Antonio Fern\'andez and V\'ictor Mart\'in-Mayor,
for letting us use the PC version of the spin-glass simulation program used in Ref.~\citenum{janus2014three}.

This work was granted access to the high performance computer (HPC)
resources of MesoPSL financed by the Region Ile de France and the
project Equip@Meso (reference ANR-10-EQPX-29-01) of the programme
Investissements d'Avenir supervised by the Agence Nationale pour la
Recherche.  This work benefited from access to the University of
Oregon HPC, Talapas, and from {\em Memento} cluster, part of the
Instituto Universitario de biocomputaci\'on y f\'isica de sistemas
complejos at the Universidad de Zaragoza (Spain).

This project has received funding from the European Research Council
(ERC) under the European Union's Horizon 2020 research and innovation
programme (grant agreement n. 723955 - GlassUniversality).  B.S. was
partially supported through Grant No. FIS2015-65078-C2-1-P, jointly
funded by MINECO (Spain) and FEDER (European Union).

\section*{Appendix}

\renewcommand\thefigure{A\arabic{figure}}
\renewcommand\theequation{A\arabic{equation}}
\setcounter{figure}{0}   
\setcounter{equation}{0}    

\appendix

\begin{figure}[t]
\centering
  \includegraphics[width=\columnwidth]{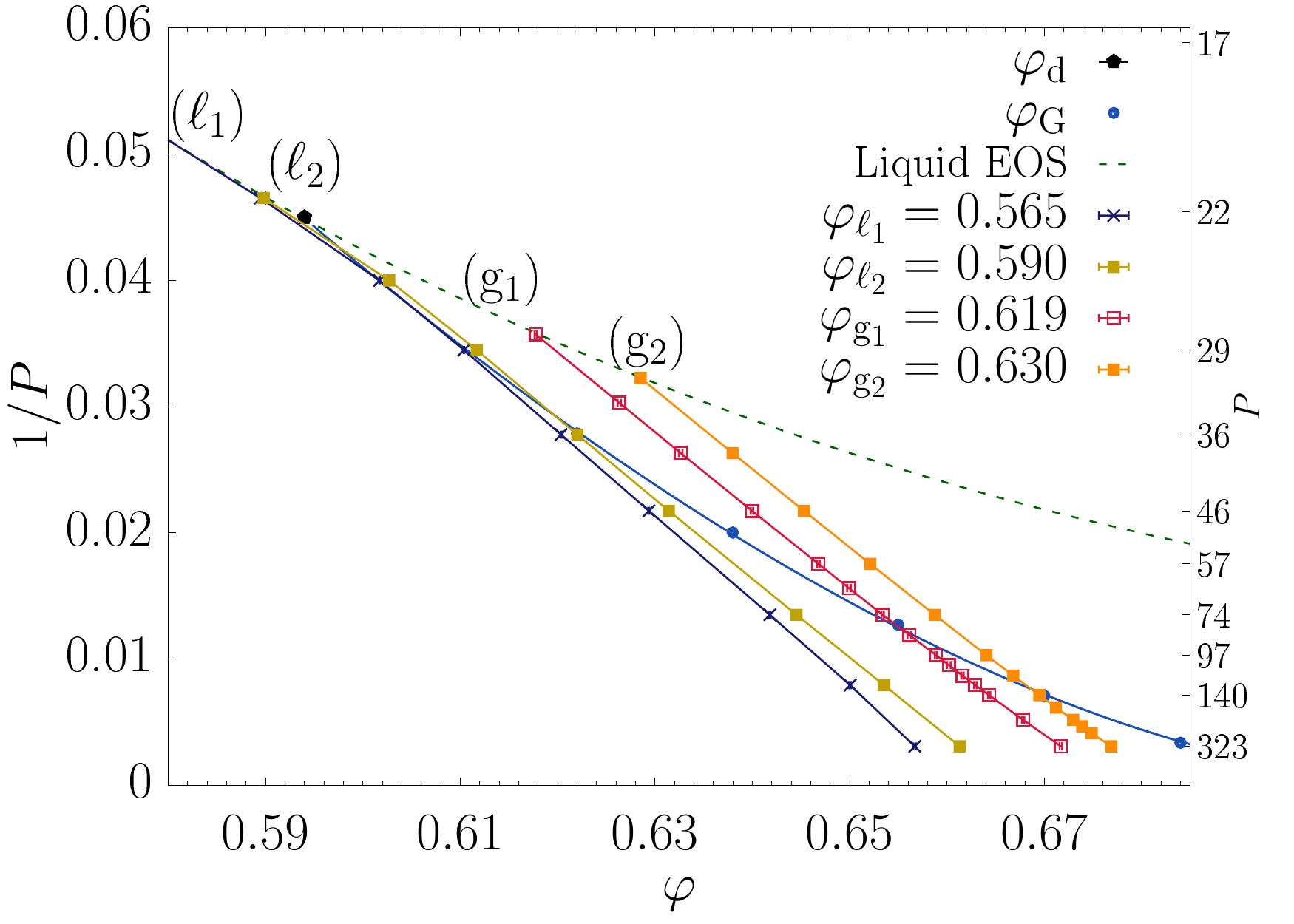}
  \caption{{\bf Two quenches from the liquid}. Same as Fig.~\ref{fgr:PD} but this time adding a second quench from the ergodic liquid phase, starting at $\varphi_{\ell_2}=0.59$ }
  \label{fgr:PD2}
\end{figure}
\begin{figure*}[t]
\centering
  \includegraphics[height=7cm]{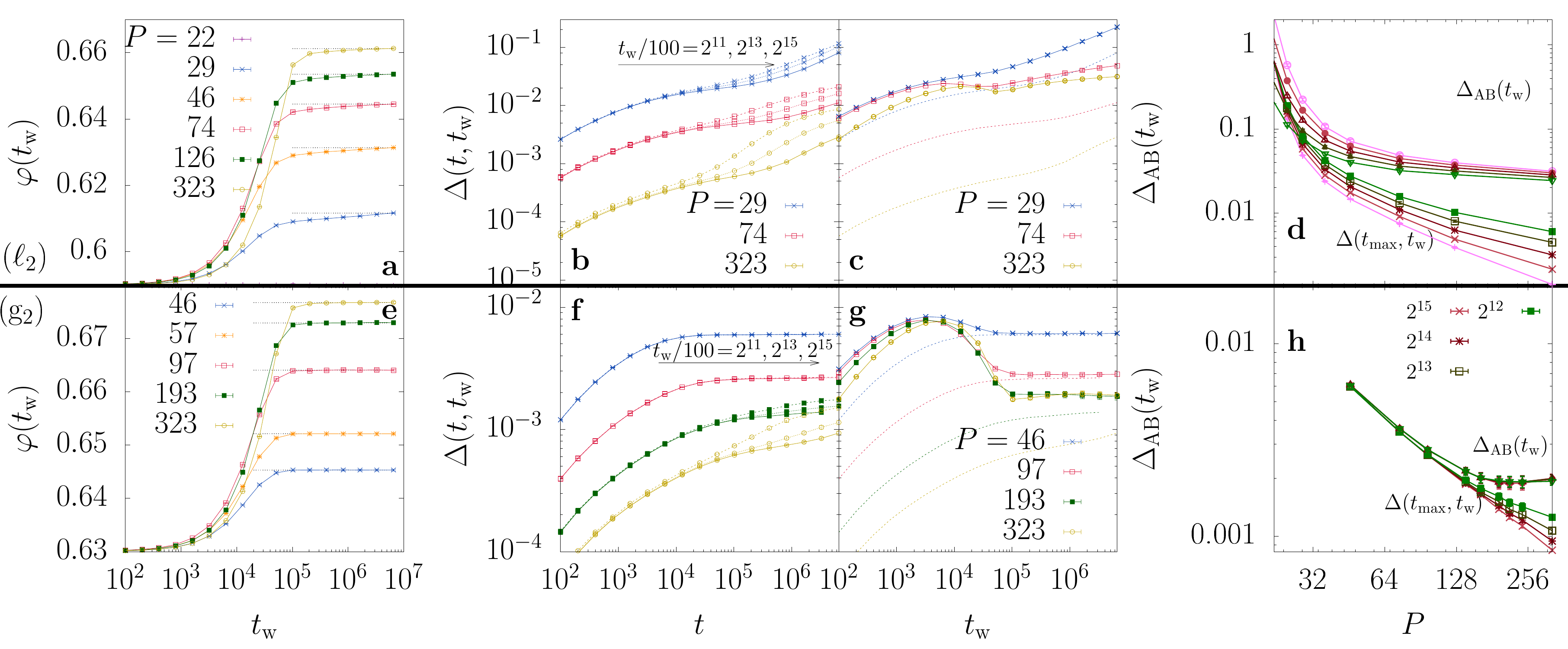}
  \caption{{\bf Dynamics after a crunch for the additional lines.} Same as Fig.~\ref{fgr:dynamics} for the two other initial states, $\varphi_{\ell_2}=0.59$ (top) and $\varphi_{\mathrm{g}_2}=0.630$ (bottom).  
    }
  \label{fgr:dynamics2}
\end{figure*}

\section{Additional results for the crunches}
In the main text we showed three off-equilibrium glass equations of state corresponding to instantaneous compressions from two dynamically arrested liquid configurations and one ergodic liquid configuration. In general, one expects that all the quenches from the ergodic liquid phase are equivalent, but this is only true if the starting point is far enough from the dynamical transition point, here $\varphi_\mathrm{d}\sim 0.594$. To study this effect, we have also repeated the same protocol described in the main text, but this time starting from equilibrium ergodic liquid configurations at $\varphi_{\ell_2}=0.59$, which is relatively near $\varphi_\mathrm{d}$. We show in Fig.~\ref{fgr:PD2} its off-equilibrium equation of state, together with the other three lines. From the data, it is clear that the new line extrapolates to a significantly higher jamming point, but also its slope becomes more similar to the one described by the two glass lines obtained from dynamically arrested configurations.
Despite this, the dynamics along this $\ell_2-$line are more similar to that of the crunches from the ergogic liquid discussed in the main text, as we show in Fig.~\ref{fgr:dynamics2}. For completeness, we also include in Fig.~\ref{fgr:dynamics2} the data for the $g_2$-line not shown in the main text. 

\section{On the number of samples and clones}

As discussed in the manuscript, we used two kinds of statistical averages: on the $N_\mathrm{c}$ clones, denoted by $\langle \bullet \rangle$, to compute the dynamical averages, and on the $N_\mathrm{s}$ samples, denoted by $\overline{\bullet}$, to compute the averages over the disorder (here initial configurations of the equilibrium liquid). In this section we discuss the dependence of the statistical errors on the susceptibilities, as function of $N_\mathrm{c}$ and $N_\mathrm{s}$, which could be useful for the design of future studies. This is particularly relevant for experimental tests, where increasing $N_\mathrm{s}$ is technically much simpler than $N_\mathrm{c}$.

\begin{figure}[t]
\centering
  \includegraphics[height=7cm]{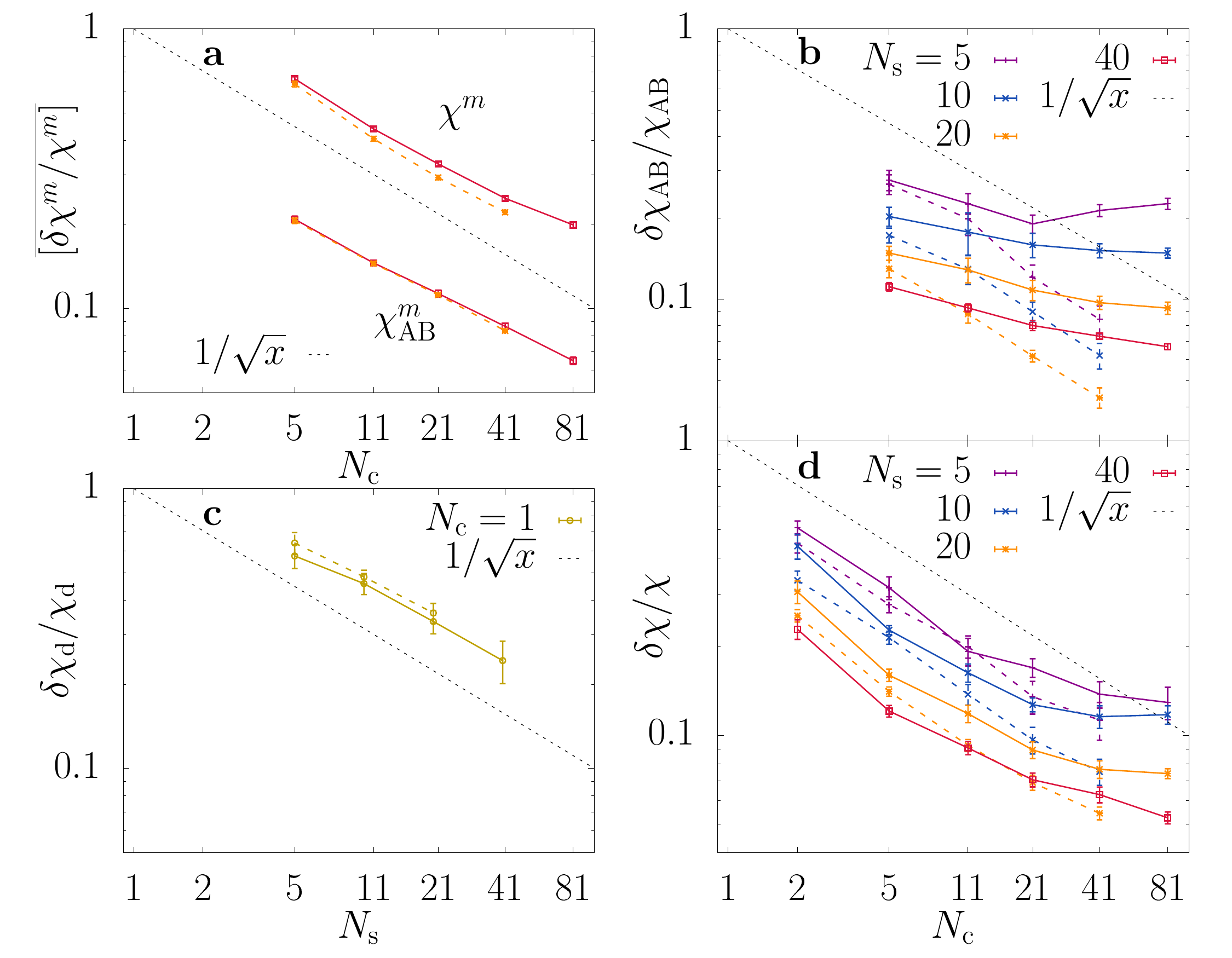}
  \caption{{\bf Relative error of the susceptibility as function of $N_\mathrm{c}$ and $N_\mathrm{s}$.} 
  Throughout the figure, we show data averaged over pressures $P>97$ on the ergodic liquid line $\varphi_\ell$ as dashed lines, 
  and over the dynamically arrested line $\varphi_{\mathrm{g}_1}$ as solid lines. In panel (a) we show the relative error over clones 
  of the sample estimate for 
  $\chi_\mathrm{AB}^m$ (defined in Eq.~\eqref{eq:chiABdefsample}) and $\chi^m$ (defined in Eq.~\eqref{eq:chidyndefsample}) averaged over all the samples. In panels (b,d), we show for different number of samples $N_\mathrm{s}$ the relative error over samples of (b) $\chi_\mathrm{AB}$ defined in Eq.~\eqref{eq:chiABdef}, and (d) $\chi$ defined in Eq.~\eqref{eq:chidyndef}. In panel (c) we show the relative error over samples for  
  $\chi_{\rm d}$ (that allows one to avoid the cloning procedure), defined in Eq.~\eqref{eq:chidyndef2}.
  }
  \label{fgr:errorsample}
\end{figure}

In the following,  the statistical error of a susceptibility $\chi$ is designated as $\delta \chi$, and  corresponds to the statistical error of the average of this magnitude over all the samples and/or clones.
In Fig.~\ref{fgr:errorsample}a we show the dependence on $N_\mathrm{c}$ of the relative statistical error of both $\chi^m_{AB}(\tw)$ and 
$\chi^m(t,\tw)$, defined in Eqs.~\eqref{eq:chiABdefsample} and ~\eqref{eq:chidyndefsample} respectively. 
For each sample, we consider the largest available times $t$ and $\tw$.
We first compute the error $\d\c^m$ on the determination of $\c^m$ for a given sample $m$ using $N_{\rm c}$ clones, 
and we then take the average $\overline{\d\c^m/\c^m}$ over the samples $m$. An additional average over all pressures $P>97$ is taken at the end,
to obtain cleaner data.
Note that the computation of $\chi^m_\mathrm{AB}$ defined in Eq.~\eqref{eq:chiABdefsample} requires at least two clones, $N_\mathrm{c}=2$. 
If $N_\mathrm{c}>2$, averages over all the possible pairs 
$N_\mathrm{c}(N_\mathrm{c}-1)/2$ of clones $A-B$ are considered. Yet, all these pairs are not statistically independent. We tackled this problem by constructing $N_\mathrm{c}$ blocks containing all the couples that contained one particular clone. The error is later extracted from the fluctuations between these blocks using the jack-knife method. 
Fig.~\ref{fgr:errorsample}a shows that, indeed, the error decays as $\EE/\sqrt{N_\mathrm{c}}$ in both cases. Yet, due to the average over many pairs of clones, the prefactor $\EE$ of $\chi_\mathrm{AB}$ is smaller than that of $\chi$, which is computed using single clones.
We conclude that, as expected, a large $N_\mathrm{c}$ allows one to obtain a perfect determination of the susceptibility $\c^m$ 
of individual samples $m$. With $N_{\rm c}=81$ clones, the statistical error is of the order of 7\% for $\chi^m_{AB}(\tw)$ and
of 20\% for $\chi^m(t,\tw)$.

Next, we discuss the sample-to-sample fluctuations. To this aim, we now consider the sample-averaged susceptibilities defined in
Eqs.~\eqref{eq:chiABdef} and \eqref{eq:chidyndef}, and their associated statistical errors assotiated to the average of $\c^m$ over samples $m$.
The relative error $\d\c / \c$ is reported in Fig.~\ref{fgr:errorsample}b for $\chi^m_{AB}(\tw)$ and in Fig.~\ref{fgr:errorsample}d 
for $\chi^m(t,\tw)$, again considering the largest available times and averaging over pressures $P>97$.
In both cases, we observe that in a crunch from the ergodic liquid phase, 
the error decays as $1/\sqrt{N_{\rm s}}$, which indicates that in this case samples and clones are equivalent. This is coherent with the discussion
of the main text.
In the dynamically arrested phase, instead, for a given $N_{\rm s}$ the error saturates upon increasing $N_{\rm c}$.
In fact, the error on individual samples becomes very small for large $N_{\rm c}$, as discussed in Fig.~\ref{fgr:errorsample}a, 
but the error remains finite due to sample-to-sample fluctuations which, 
as expected, are present in this case (as discussed in the main text around Fig.~\ref{fgr:susc-samples}). Upon increasing the number of 
samples $N_{\rm s}$, the error decays as $1/\sqrt{N_{\rm s}}$, as expected.

Finally, we discuss the determination of $\chi$ in the case where no clone can be constructed, as it would be most often the case in experiments. 
Note that in this case, the definition of Eq.~\eqref{eq:chidyndef} is not longer valid, as in presence of a single clone one would have $\c(t,\tw)=0$ by
construction. A slight variation must then be considered
\begin{equation}\label{eq:chidyndef2} 
\chi_\mathrm{d}(t,\tw) =\frac{
      \sum_{ij} [ \overline{ u_i(t,\tw) u_j(t,\tw)} -
      \overline{u_i(t,\tw)}\,\,
      \overline{u_j(t,\tw)} ] }{ \sum_i [\overline{ u_i(t,\tw)^2} -
      \overline{ u_i(t,\tw)^2}]}  \ ,
      \end{equation}
      where here the average is over the $N_{\rm s}$ samples, with a single clone per sample.
We checked that, in our simulations, the differences between $\chi$ and $\chi_\mathrm{d}$ are within the errorbars of $\chi$. 
In Fig.~\ref{fgr:errorsample}c we show the relative statistical error on the determination of $\c_{\rm d}$ which, as expected, decays
as $1/\sqrt{N_{\rm s}}$.

\clearpage



\balance


\bibliography{HS,library} 
\bibliographystyle{rsc} 

\end{document}